\documentclass[preprint,review,12pt]{elsarticle}
\usepackage{float}
\include{packages}

\def\nobreakbefore{%
	\relax\ifvmode\else
	\ifhmode
	\ifdim\lastskip > 0pt\relax
	\unskip\nobreakspace
	\fi
	\fi
	\fi
}
\let\oldcite\cite
\renewcommand\cite{\nobreakbefore\oldcite}
\newcommand{\etal}{et al.~}

\newcommand{\EOS}{EOS\xspace}

\newcommand{\LS}{LS\xspace}

\newcommand{\DG}{DG\xspace}
\newcommand{\DGSEM}{DGSEM\xspace}
\newcommand{\FV}{FV\xspace}

\newcommand{\RK}{RK\xspace}

\newcommand{\HLLC}{HLLC\xspace}

\newcommand{\pderivative}[2]{\frac{\partial #1}{\partial #2}} %

\newcommand{\kron}[2]{\delta_{#1 #2}}

\newcommand{\gradient}[1]{\nabla_{#1}}

\renewcommand{\vec}[1]{\mathbf{#1}}

\newcommand{\zeit}{t}

\newcommand{\elem}[1]{C_{#1}} %

\newcommand{\xvec}{\vec{x}} %
\newcommand{\xt}{\xvec,\zeit} %

\newcommand{\wvec}{\vec{W}} %
\newcommand{\uvec}{\vec{U}} %

\newcommand{\fvec}{\vec{F}} %
\newcommand{\ghostflux}{\vec{G}} %

\newcommand{\heatflux}[1]{q_{#1}}
\newcommand{\stress}{\boldsymbol{\tau}}
\newcommand{\viscosity}{\mu}
\newcommand{\conductivity}{\lambda}

\newcommand{\angstrom}{\mathring{A}}

\newcommand{\cv}{c_{\specvol}}

\newcommand{\relaxNewton}[1]{z_{\text{N}}}

\newcommand{\dens}{\rho}
\newcommand{\specvol}{\nu}
\newcommand{\momvec}{\vec{m}}
\newcommand{\mom}[1]{m_{#1}}
\newcommand{\velvec}{\vec{v}}
\newcommand{\vel}[1]{v_{#1}}
\newcommand{\pres}{p}
\newcommand{\temperature}{T}

\newcommand{\Etot}{E}
\newcommand{\etot}{e}
\newcommand{\einner}{\epsilon}

\newcommand{\levelset}{\phi}

\newcommand{\levelsetvel}[1]{{v}_{#1}^{\text{\tiny\LS}}}

\usepackage{siunitx}   %
\DeclareSIUnit{\sidens}{\kilo\gram\per\cubic\meter}
\DeclareSIUnit{\sivel}{\meter\per\second}
\DeclareSIUnit{\sisigma}{\joule\per\square\meter}
\DeclareSIUnit{\sicvcp}{\joule\per\kelvin\per\kilo\gram}
\DeclareSIUnit{\sipres}{\pascal}
\DeclareSIUnit{\sipresa}{\bar}
\DeclareSIUnit{\sitemp}{\kelvin}
\DeclareSIUnit{\sizeit}{\second}
\DeclareSIUnit{\silen}{\meter}
\DeclareSIUnit{\sij}{\kelvin\square\meter\second\per\kilo\gram}
\DeclareSIUnit{\simolarflux}{\mol\per\square\meter\per\second}
\DeclareSIUnit{\simolargibbs}{\joule\per\mol}
\DeclareSIUnit{\siatomicmass}{\atomicmass}
\DeclareSIUnit{\siangstrom}{\angstrom}

\journal{Journal of Computational Physics}

\begin{document}

\begin{frontmatter}

\title{A Viscous and Heat Conducting Ghost Fluid Method for Multi-Fluid Simulations}

\author[label1]{Steven J\"ons}
\ead{steven.joens@iag.uni-stuttgart.de}
\author[label1]{Christoph M\"uller}
\author[label1]{Johanna Hintz}
\author[label1]{Andrea Beck}
\author[label1]{Claus-Dieter Munz}
\address[label1]{Institute of Aerodynamics and Gas Dynamics, University of Stuttgart, 70569 Stuttgart, Germany}

\begin{abstract}
The ghost fluid method allows a propagating interface to remain sharp during a numerical simulation. The solution of the Riemann problem at the interface provides
proper information to determine interfacial fluxes as well as the velocity of the phase boundary. When considering two-material problems, the initial states of
the Riemann problem belong to different fluids, which may have different equations of states. In the inviscid case, the solution of the multi-fluid Riemann
problem is an extension of the classical Riemann problem for a single fluid. The jump of the initial states between  different fluids generates waves in both
fluids and induces a movement of the interface, similar to a contact discontinuity. More subtle is the extension to viscous and heat conduction terms which is
the main foucs of this paper. We account for the discontinuous coefficients of viscosity and heat conduction at the multi-fluid interface and derive solutions of
the Riemann problem for these parabolic terms, from which we can derive parabolic, interfacial fluxes. We demonstrate the accuracy, robustness and applicability
of this approach through a variety of test cases.
\end{abstract}


\end{frontmatter}

\begin{section}{Introduction}
\label{sec:introduction}
In the numerical computation of multi-material and multiphase flows, a great challenge is to model the behavior of the  interfaces correctly. This includes for many technical
applications beside the movement of the interfaces also surface tension, viscous effects and heat conduction.
All these interfacial phenomena may strongly influence the fluid flow in large parts of the surrounding region. Hence, key
aspect of any numerical two-phase method is the correct balancing of all interfacial forces and fluxes such that it accurately predicts the physics. We restrict ourselves in the following to two fluids or two phases of a fluid without taking into account phase transitions. For simplicity, we name the two different fluids liquid and gas, which is our main field of application.

In general, one can distinguish between two different classes of numerical methods for two-material or two-phase flows within a continuum description:
diffuse-interface (DI) and sharp-interface (SI) methods.  In the DI methods, the interface in the macroscopic formulation is assumed to have a finite thickness.
Specifying a continuous  variation of the states in a mixing zone,
the interfacial region is smeared out over a couple of grid cells, which then allows an application of a standard continuum approximation method. In this case, the
inclusion of viscous and thermal terms, which depend on the gradients, is in principle straightforward. However, since the thermodynamic properties at the
different sides of the interface may strongly differ and the separate states may obey different equations of state, it is often difficult to establish the
thermodynamic consistency of the mixture state and the proper behavior in the vicinity of the interface. Additionally, heat conduction and viscous effects may
be difficult to approximate in a physical consistent way because of the artificially diffused interface. In physics the width of the transition zone may be in
the order of a few molecules and far away from its approximate counterpart.

The SI class of methods assumes that the material interface between the fluids is infinitely thin on the macroscopic scale, which results in a discontinuity
within a macroscopic continuum description. A method to keep a propagating interface sharp is to move the grid according to the interface motion. Other
sharp-interface methods rely on an interface reconstruction, cut-cells or the  ghost fluid method (\cite{Fedkiw1999}). In the latter cases, the solution of the
flow equations have to be combined with a tracking of the interface movement.  Several successful methods have been developed in the past including the
volume-of-fluid (\cite{Hirt1981}), the front-tracking (\cite{Tryggvason2001}), and the level-set methods (\cite{Sussman1994}).
Beside the interface tracking, the imposing of coupling conditions at the interface is another building block in the sharp-interface approach. This coupling consists of a set of jump conditions for the macroscopic variables, modelling the local physical behavior. Jump conditions may be obtained from the solution of the Riemann problem for multi-material or multiphase flow.

Due to the discontinuous nature of the physical quantities at the interface in the SI approach, the inclusion of viscous and thermal effects are challenging.
This topic was addressed by Fedkiw \etal in \cite{Fedkiw2001}  by extending the ghost fluid methodology to the viscous stress tensor. The artificial ghost states are derived from a splitting of the viscous tensor in continuous and discontinuous terms. The gradients across the interface are evaluated by assuming a continuous transition, while the tangential terms are approximated in an upwind fashion. Another approach is the continuous surface force method, which approximates the parabolic terms by imposing continuity and regularizing the $\delta$-function in a small neighborhood (\cite{Brackbill1992, Sussman1994}). Hence, the viscous terms are reformulated from a surface force to a volume force as was done for the surface tension force in \cite{Sussman1994}. This
new source term is only non-zero in a narrow band around the interface and smears the originally singular force out in this area.
Both methods rely on a decomposition of the viscous stress tensor in a continuous and discontinuous part with a standard finite difference approximation for the
continuous gradient calculation and with some complex workaround for the discontinuous part.

The approach under consideration in this paper is motivated by the use of the Riemann problem in the  ghost fluid method for the hyperbolic terms as proposed in \cite{Liu2003, Merkle2007, Fechter2018}. We extend the diffusive Riemann solver of Gassner \etal \cite{Gassner2007a, Gassner2007} to the two-fluid case and propose a thermal and a viscous
Riemann solver. According to Godunov's
idea for hyperbolic conservation equations, the local solution is used to determine a numerical flux. Gassner \etal derived in \cite{Gassner2007a} an exact Riemann solver for the scalar, linear diffusive generalized Riemann problem and introduced a linearization for the nonlinear case. In \cite{Gassner2007}, this methodology was successfully applied
to the compressible Navier-Stokes equations. In \cite{Loercher2008} it was additionally analyzed in combination with discontinuous Galerkin schemes. When using this approach,
no preliminary deconstruction of the viscous stress tensor nor a continuity assumption approximating the gradient across the interface is needed. The Riemann
solvers are validated as building blocks in a sharp-interface level-set ghost fluid method to calculate interfacial diffusive fluxes.

As already mentioned, we restrict ourselves to the two-fluid case without phase transition. For simplicity, we name the different phases liquid and gas.
The paper has a simple structure as follows: In Section \ref{sec:method}, we discuss the governing equations and all the basic building blocks of our numerical approach.
Afterwards, we describe our modification of the ghost fluid method in section \ref{sec:gfm}. Here, we define  Riemann solvers for heat conduction and
friction.  In section \ref{sec:results} we validate our  method and apply it to test cases. Finally, we conclude our results
in section \ref{sec:conclusion}.
\end{section}

\begin{section}{Fundamentals} \label{sec:method}
In the present section, the governing equations and the fundamental building blocks of the employed numerical scheme are introduced.
\begin{subsection}{Governing Equations}
\label{sec:mathematical_model}
We consider compressible,  viscous and heat conducting two fluid flows without phase transition. The fluid flow in each phase can be described by the Navier-Stokes equations (NSE):
 \begin{align}
 \label{eq::nse}
 \pderivative{\uvec}{\zeit} + \sum\limits_{i=1}^{3} \pderivative{}{x_{i}}
 \left[
   \fvec_{h}^{i}
 - \fvec_{p}^{i}
 \right]= 0
 \end{align}
 with the state vector
 \begin{align}
  \label{eq::qvec}
  \uvec = \uvec(\xvec,\zeit) =
  \begin{pmatrix}
    \dens \\
    \mom{1} \\
    \mom{2} \\
    \mom{3} \\
    \Etot
  \end{pmatrix}
  =
  \begin{pmatrix}
    \dens \\
    \dens \vel{1} \\
    \dens \vel{2} \\
    \dens \vel{3} \\
    \dens \etot
  \end{pmatrix},
 \end{align}
 the hyperbolic flux
 \begin{align}
  \fvec_{h}^{i} = \fvec_{h}^{i}(\uvec)  =
  \begin{pmatrix}
    \dens         \vel{i} \\
    \dens \vel{1} \vel{i} + \kron{1}{i} \pres\\
    \dens \vel{2} \vel{i} + \kron{2}{i} \pres\\
    \dens \vel{3} \vel{i} + \kron{3}{i} \pres\\
     (\dens\etot + \pres)   \vel{i}
  \end{pmatrix}
\end{align}
and the parabolic flux
\begin{align}
  \fvec_{p}^{i} = \fvec_{p}^{i}(\uvec, \gradient{} \uvec)  =
  \begin{pmatrix}
    0 \\
    \tau_{1i}\\
    \tau_{2i}\\
    \tau_{3i}\\
    \sum\limits_{j=1}^{3}\tau_{ij} \vel{j} -\heatflux{i}
  \end{pmatrix}.
\end{align}
Thereby, the density $\dens$, the momentum $\momvec = (\mom{1},\mom{2},\mom{3})^T$ and the total energy per unit volume $\Etot$ are the conserved variables. Additional quantities are the velocity $\velvec = \momvec /\dens$, the total energy per unit mass $\etot = \Etot / \dens$ and the pressure $\pres$. Furthermore, the viscous stress tensor is defined as
\begin{align}
  \underline{{\stress}} = \viscosity   \left( \nabla \velvec + \left(\nabla \velvec\right)^T - \dfrac{2}{3} (\nabla \cdot \velvec) \underline{\underline{I}} \right)
\end{align}
and the heat flux is given by
\begin{align}
  \vec{\heatflux{}} =
  \begin{pmatrix}
    \heatflux{1}\\
    \heatflux{2}\\
    \heatflux{3}
  \end{pmatrix}
  = -\conductivity \nabla \temperature
\end{align}
with the dynamic viscosity $\viscosity = \viscosity(\dens,\temperature)$, the thermal conductivity $\conductivity=\conductivity(\dens,\temperature)$ and the temperature $T$.
The total energy is the sum of internal and kinetic energy
\begin{align}
  \label{eq::etot}
  \Etot = \dens \etot = \dens \einner +\dfrac{1}{2}\dens \velvec\velvec^T
\end{align}
with the internal energy per unit mass $\einner$. The equation system (\ref{eq::nse}) is closed by an  equation of state (\EOS):
\begin{align}
  \einner = \einner (\dens,\temperature), \\
  \pres   = \pres (\dens,\temperature)
\end{align}
that relates the internal energy per unit mass and the pressure to density and temperature.
\end{subsection}

\begin{subsection}{Sharp-Interface Method}
   In the following, we give a short overview of the employed sharp-interface method. It is based upon three basic building blocks: The bulk phase flow solver,
   the interface capturing and the ghost fluid method. A more detailed discussion can be found in \cite{joens2020}. We discretize
   the computational domain into $\elem{n}$ hexahedral grid cells. The fluid flow in the bulk, described by the NSE, is solved with the discontinuous Galerkin
   spectral element  method (\DGSEM), detailed in \citet{Hindenlang2012} and \citet{Krais2020}.  The \DGSEM is a high-order method with a  tensor-product based polynomial solution
   representation of degree $N$ in each grid element. The individual elements are coupled via Riemann solvers, similar to the well-known finite volume (FV)
   methods. The \DGSEM is supplemented by the second order FV shock capturing scheme \cite{Sonntag2016} which reinterprets a grid element into a
   set of $(N+1)^d$ equidistantly spaced FV sub-cells. In this way strong gradients, e.g., shock waves,  in the solution can be captured,  avoiding spurious
   oscillations of the DG polynomial. Solution gradients for the parabolic terms are obtained via the BR1-lifting procedure of \citet{Bassi1997}. Time integration is performed using an explicit fourth order low storage \RK scheme \cite{Kennedy2003}.\\

   The position of the interface is tracked via the level-set method of \citet{Sussman1994}. Therein, the interface is defined as the zero of the level-set
   function $ \levelset(\xt)$, which is a signed distance function. Advection of the interface is treated via an additional advection equation, the level-set
   transport equation
  \begin{align}
    \label{eq:lstransport}
    \pderivative{\levelset(\xt)}{\zeit} + \sum_{i=1}^{3}\levelsetvel{i}(\xt)\pderivative{\levelset(\xt)}{x_i} = 0,
  \end{align}
  with $\levelsetvel{i}$ being the level-set transport velocity in  direction $i$.
  The numerical treatment of Eq. \eqref{eq:lstransport} follows \citet{joens2020}, in using the path-conservative approach of Dumbser and Loub\`ere
   \cite{Dumbser2016a} within the DGSEM with FV shock-capturing. In this way, also the level-set field is obtained with high-order accuracy. From
   $\levelset(\xt)$, geometrical properties of the interface as normal vectors can be calculated, as discussed in \cite{joens2020,Fechter2015a}. The advection of the level-set
   field does not conserve the signed-distance property, therefore it is enforced additionally via a reinitialization procedure. Here we follow the approach of
   \citet{Peng1999} in using a 5th-order WENO scheme to solve a Hamilton-Jacobi type equation. A similar procedure is used to obtain the level-set advection velocity
   \cite{Peng1999}, which is initially only known at the interface and needs to be extrapolated into the volume.

   Finally, the ghost fluid Method (GFM) \cite{Fedkiw1999} is used to couple the individual phases with each other. In order to apply the GFM, we employ a
   domain decomposition based on the sign of the level-set function which identifies a grid element as liquid or vapor. In grid elements, where an interface is
   present, the switch to FV sub-cells is applied and the phase of each sub-cell is identified as shown in Fig.
   \ref{fig:levelsetghostfluid_domaindecomposition}. This procedure thus defines a numerical interface, denoted as the surrogate phase interface, which is the
   discrete representation of the level-set function and coincides with the sub-cell sides. In the ghost fluid method, ghost states or ghost fluxes are
   introduced across these interfacial sides as boundary conditions for the bulk solver in the gas as well as in the liquid. In this way, time evolution of each phase becomes trivial
   if the states of the ghost fluid are known. These need to be defined in such a manner that they model the interaction at the multi-fluid interface
   appropriately.  For the definition of the ghost state, we do not follow the original variant of \citet{Fedkiw1999} but employ the Riemann solver technique as
   introduced in \citet{Liu2003, Merkle2007, Fechter2018}. For the multi-fluid case, the GFM based on solution of Riemann problems at the interface have been
   considered in \citet{Xu2011,Xu2016,Fechter2015}. The solution of the Riemann problem is used to
   define the ghost states as well as the level-set transport velocity at the interface.

  \begin{figure}[h]
    \centering
    \includestandalone[width=0.7\textwidth]{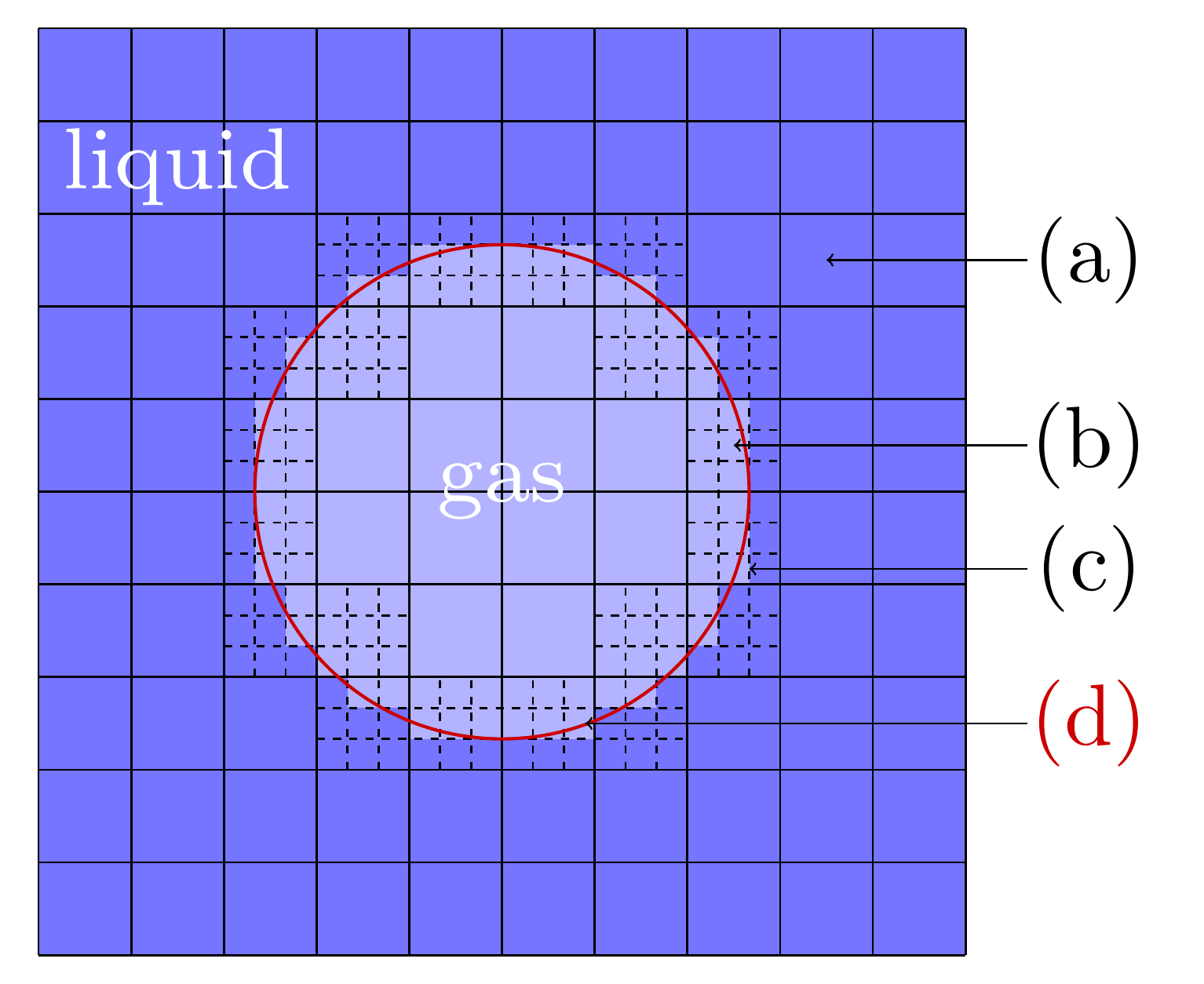}
    \caption{2D domain decomposition: \DG elements (a), \FV sub-cells (b), the surrogate phase boundary (c) and the actual phase boundary (d), which is identified by the root of the level-set function, are shown.}
    \label{fig:levelsetghostfluid_domaindecomposition}
  \end{figure}
\end{subsection}
\end{section}

\begin{section}{A Hyperbolic-Parabolic Extension of the Ghost Fluid Method}
	\label{sec:gfm}
   The ghost fluid method, in which the ghost states are calculated by solving Riemann problems, has originally been considered for purely hyperbolic equation
   systems. In the present section, we extend it to the Navier-Stokes equations. We note that this procedure is more general and may be also applied to other
   hyperbolic-parabolic equation systems. Instead of ghost states, from which the numerical fluxes are calculated, we formulate the method by directly defining
   interfacial ghost fluxes as used in \citet{Fechter2018}. Note that, when employing an exact Riemann solver in the hyperbolic case, both approaches are identical.
   The interfacial fluxes are written as the sum of the hyperbolic, the viscous and the thermal flux:
   \begin{align}
  	\fvec_{}^{*} = \fvec_{Euler}^{*}+\fvec_{Therm}^{*} +\fvec_{Visc}^{*}.
  \end{align}
  As usual in the single fluid case, the three contributions are calculated separately from each other. All interfacial fluxes are calculated by employing
  appropriate Riemann solvers. Before solving the parabolic Riemann problems we need an additional step.
  The parabolic fluxes depend on gradients at the interface. Hence, beside the states we need the gradients from the liquid and from the vapor region.
  On each side of the interface, the gradient is calculated based on solely the information from the same phase, allowing a jump in the gradients.
  We calculate these gradients by a masked least squares method, based on the approach of \cite{foll2020novel}. All entries in the least squares
  matrix pertaining to the opposite phase are set to zero and a one-sided gradient is calculated. These gradients are then further used as initial conditions
  for the Riemann problem or generalized Riemann problem as often called in the hyperbolic case for piecewise linear initial data. As usual, we use the rotational invariance of the Navier-Stokes
  equations, which allows us to consider one-dimensional Riemann problems into the normal direction $\xi$ with respect to the surrogate phase boundary. We rotate the
  states in such a way that the left state is always liquid and the right state is always gas.

  The main novelty and hence, the focus of this work, lies in the use of Riemann problems for heat conduction and friction at the interface. The use of Riemann problems for parabolic
  terms, similar to it's hyperbolic counterpart in classical FV methods, was introduced by \citet{Gassner2007a} in the form of the diffusive generalized Riemann
  problem (dGRP) in the scalar case. The authors solved a piecewise polynomial Riemann problem exactly and used this time dependent exact solution to define a numerical diffusion flux. Follow up
  investigations of \citet{Loercher2008} and \citet{Gassner2007} considered heat conduction with varying diffusion coefficients as well as the application to the
  NSE. Given the differences in density across the interface, the mentioned approaches can not be straightforwardly applied to the sharp-interface situation
  considered in this paper. Hence, we extend the work from \cite{Gassner2007,Gassner2007a,Loercher2008} in the following. We begin by considering a general
  scalar diffusion equation with jumping diffusion coefficients and derive a solution for this Riemann problem. This general solution is then used to obtain solutions
  of the thermal and viscous Riemann problem and fluxes for the ghost fluid method. We shortly describe the hyperbolic flux for
  completeness.

  \begin{subsection}{The Interfacial Riemann Solver for the Hyperbolic Terms}
  \label{sec::ghostfluxes}
  The hyperbolic Riemann problem is defined by the Euler equations
  \begin{equation}
     \uvec_t+ \fvec_{Euler}(\uvec)_{\xi}=0
     \label{eq:euler}
  \end{equation}
  with the piecewise constant initial data
  \begin{equation}
     \uvec(\xi,t) =
     \begin{cases}
        \uvec_{liq} \quad &\text{for} \quad \xi<0,\\
        \uvec_{vap} \quad &\text{for} \quad \xi\ge0,
     \end{cases}
     \label{eq:eulerinitial}
  \end{equation}
  where $\xi$ denotes the direction normal to the interface. The solution to this problem is obtained via the variant of the \HLLC Riemann solver of Toro
  \cite{Toro1994} for fluid interfaces, which was proposed by Hu \etal \cite{Hu2009}. The assumed wave pattern for this solution is depicted in Fig.
  \ref{fig:riemann_euler_hllc}. It consists of two outer waves and a material interface that separates the two inner states.  Surface tension may be included
  similar to Jaegle \etal \cite{Jaegle2012} at the intermediate wave in form of the Young-Laplace Law.
    \begin{figure}[h]
       \centering
       \includestandalone[width=.8\textwidth]{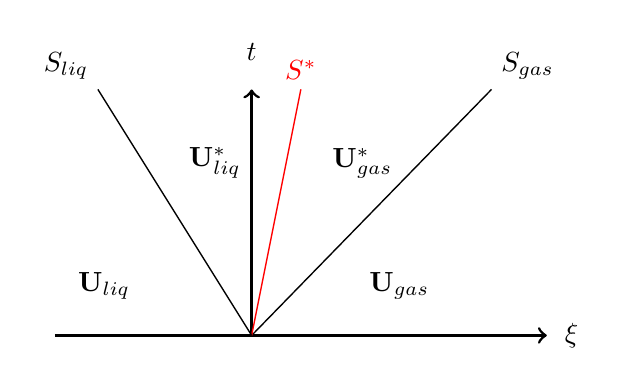}
         \caption{Approximate hyperbolic two-phase Riemann solution (HLLC).}
         \label{fig:riemann_euler_hllc}
     \end{figure}

 \end{subsection}

  \begin{subsection}{The Generalized Scalar Riemann Problem for Diffusion with Discontinuous Coefficients}
  \label{sec::dgrp_diff}
   We first consider a scalar, one-dimensional diffusion equation
    \begin{align}
      \label{eq::diff}
      u_t - \left(a u_{\xi}\right)_{\xi}=0
    \end{align}
    with the diffusion coefficient $a$. We want to solve the generalized Riemann problem with piecewise linear initial data
    \begin{align}
      \label{eq::dgrp}
     u(\xi,\zeit=0)=
      \begin{cases}
         u^{-}+\xi  u_{\xi}^- \quad \text{for} \quad \xi<0,\\
         u^{+}+\xi  u_{\xi}^+ \quad \text{for} \quad \xi\geq0,
      \end{cases}
    \end{align}
    with the positive and piecewise constant diffusion coefficient
    \begin{align}
      \label{eq:ICa}
       a(\xi)=
       \begin{cases}
          a^- \quad \text{for} \quad \xi<0, \\
          a^+ \quad \text{for} \quad \xi\geq0.
       \end{cases}
    \end{align}
    To solve this initial value problem, we follow \citet{Loercher2008} and apply the Laplace transformation  to equation \eqref{eq::diff}. Denoting $\Theta$ as the Laplace transform of $u$
    \begin{align}
      \label{eq::Theta}
      \mathcal{L}\{u(\xi,\zeit)\}=\Theta(\xi,s),
    \end{align}
    the transformation of \eqref{eq::diff} reads as
    \begin{align}
       \partial_{\xi}^2\Theta - \frac{s}{a} \Theta =-  \frac{u^- +  \xi u_{\xi}^-}{a}.
       \label{eq:laplace}
    \end{align}
  The solutions to this ordinary differential equation is given by
  \begin{align}
     \Theta& = c \exp \left( \sqrt{\frac{s}{a}}\xi \right) + \frac{u^-}{s} + \frac{u_{\xi}^-}{s}\xi  \quad \text{for} \quad \xi<0,
  \label{eq:arbExactsol}
  \end{align}
  with $c=c(\xi)$ being constant at each side of the interface. The values are obtained by applying compatibility conditions at
  the interface, $\xi=0$:
  \begin{align}
  \label{eq::contact}
  \Theta^-(0,s)&=\Theta^+(0,s),\\
  \label{eq::contact2}
  b^- \pderivative{}{\xi} \Theta^-(0,s)&=b^+\pderivative{}{\xi} \Theta^+(0,s).
\end{align}
The first condition imposes the continuity of the solution. The second condition imposes the continuity of a flux. This condition is a more general
than the one used by \citet{Loercher2008}, where solely the continuous diffusion flux with $b^-=a^-$ and $b^+=a^+$ was imposed. This allows us
to extend this solution to the multi-fluid situation, as will be seen later.  From the two conditions \cref{eq::contact,eq::contact2}, the coefficients $c^-$
and $c^+$ for $x<0$ and $x\ge0$, respectively, can be determined as
\begin{subequations}
  \begin{align}
     c^-=\frac{ b^+ \sqrt{a^-} (u^+-u^-) + \sqrt{\frac{a^+a^-}{s}}(b^+u^+_{\xi}-b^- u^-_\xi)} {s(b^+\sqrt{a^-}+b^-\sqrt{a^+})},  \\
     c^+=\frac{-b^- \sqrt{a^+} (u^+-u^-) + \sqrt{\frac{a^+a^-}{s}}(b^+u^+_{\xi}-b^-u^-_\xi)}  {s(b^+\sqrt{a^-}+b^-\sqrt{a^+})},
  \end{align}
  \label{eq:coefficients}
\end{subequations}
giving the solution of the dGRP. To calculate the numerical flux from this solution, we need the derivatives of $\Theta$ from \cref{eq:arbExactsol} with respect to $\xi$:
\begin{subequations}
   \begin{align}
      &\left(\pderivative{}{\xi} \Theta^-\right) (\xi,s)= c^-\sqrt{\frac{s}{a^-}} \exp \left(  \sqrt{\frac{s}{a^-}} \xi \right) + \frac{u^-_\xi}{s}, \\
      &\left(\pderivative{}{\xi} \Theta^+\right) (\xi,s)=-c^+\sqrt{\frac{s}{a^+}} \exp \left( -\sqrt{\frac{s}{a^+}} \xi \right) + \frac{u^+_\xi}{s}.
   \end{align}
\label{eq:derrivarbsolution}
\end{subequations}
Applying the inverse Laplacian, we obtain the physical derivatives as
   \begin{align}
   \pderivative{}{\xi} u^-(0,t)=\frac{b^+ (u^+-u^-)}{\sqrt{\pi t}(b^+\sqrt{a^-}+b^-\sqrt{a^+})} + \frac{\sqrt{a^+}b^+u^+_\xi +  b^+\sqrt{a^-} u^-_\xi} {b^+\sqrt{a^-}+b^-\sqrt{a^+}}, \\
   \pderivative{}{\xi} u^+(0,t)=\frac{b^-(u^+-u^-)}{\sqrt{\pi t}(b^+\sqrt{a^-}+b^-\sqrt{a^+})} + \frac{\sqrt{a^+}b^-u^+_\xi + b^-\sqrt{a^-} u^-_\xi}{b^+\sqrt{a^+}+b^-\sqrt{a^+}}.
\label{eq:x0derrivative}
 \end{align}
From  these solution gradients, we obtain the numerical diffusion flux by taking the integral mean in time for one time step
\begin{equation}
   \ghostflux = \frac{1}{\Delta t} \int_0^{\Delta t}b\pderivative{u}{\xi}(0,t) \mathrm{d}t.
\end{equation}
We note that the derivatives are singular at $\xi=0$ as derivatives of a discontinuous function. However, the time average exists as an improper integral and reads as
\begin{align}
\ghostflux &=\frac{2 b^+ b^- (u^+-u^-)}{\sqrt{\pi \Delta t}(b^+\sqrt{a^-}+b^-\sqrt{a^+})} + \frac{ b^+ b^- (\sqrt{a^+} u^+_{\xi} + \sqrt{a^-} u^-_{\xi})}{(b^+\sqrt{a^-}+b^-\sqrt{a^+})}.
\label{eq:intheatflux}
\end{align}
From Eq. \eqref{eq:intheatflux}, the case studied by L\"orcher \etal\cite{Loercher2008} is recovered for $b^- = a^-$ and $b^+ = a^+$.
\end{subsection}

\begin{subsection}{The Numerical Heat Flux}
 \label{sec::thermalsub}
 In the following, we want to derive a formulation for the numerical heat flux. Therefore, we want to solve the thermal Riemann problem normal to the interface defined by
 \begin{equation}
    \uvec_t -\fvec_{Therm}(\uvec)_\xi=0,
    \label{eq:thermRP}
 \end{equation}
 with the thermal flux $\fvec_{Therm}(\uvec)=(0,0,0,-q_1)^{\mathrm{T}}$, and the piecewise linear initial data
\begin{align}
   \label{eq:grp}
   \uvec(\xi,\zeit=0)=
   \begin{cases}
      \uvec_{liq}^-+\xi \left(\uvec_{liq}\right)^-_\xi  \quad &\text{for} \quad \xi<0\\
      \uvec_{gas}^++\xi \left(\uvec_{gas}\right) ^+_\xi \quad &\text{for} \quad \xi\geq0
   \end{cases}.
\end{align}
We further assume both time independence and a piecewise constant distribution of the thermal conductivity $\lambda$ and the specific heat at constant volume
$c_v$. Eq.\eqref{eq:thermRP} is identical to considering the energy equation
\begin{align}
    \label{eq:thermsuba}
    \dfrac{\partial}{\partial \zeit}\left( \dens\einner\right)_t  +\dfrac{\partial}{\partial \xi} \left(\conductivity \dfrac{\partial}{\partial \xi} \temperature\right)  &= 0
 \end{align}
under the condition of a time independent density and momentum. In the following, we reformulate Eq. \eqref{eq:thermsuba} to an equation for the temperature in
order to receive a connection with the numerical flux constructed in the previous subsection.  We utilize the thermodynamic identity
 \begin{align}
   \label{eq::deps}
   \mathrm{d}\einner &= \left(\frac{\partial \einner}{\partial \temperature}\right)_{\dens} \mathrm{d} \temperature  +  \left(\frac{\partial \einner}{\partial \dens}\right)_{\temperature} \mathrm{d}\dens \\
   &= \cv \mathrm{d} \temperature  +  \left(\frac{\partial \einner}{\partial \dens}\right)_{\temperature} \mathrm{d}\dens,
\end{align}
and obtain
\begin{align}
 \dfrac{\partial}{\partial \zeit} \left(\dens \cv \temperature\right) + \dfrac{\partial}{\partial \xi} \left(\conductivity \dfrac{\partial}{\partial \xi} \temperature\right)  &= 0.
\end{align}
With the  assumptions on $\lambda$ and $c_v$ we can formulate the temperature equation as
\begin{align}
   \pderivative{\temperature}{\zeit}  + k \dfrac{\partial^2}{\partial \xi^2} \temperature &= 0,
\end{align}
with the thermal diffusivity
   \begin{align}
     k = \dfrac{\conductivity}{\dens \cv}.
   \end{align}
Reinterpreting the initial data to
\begin{align}
  \temperature(\xi,0)=
  \begin{cases}
     \temperature_{liq}+\xi(\temperature_{liq})_\xi \quad & \text{and } \conductivity_{liq} = const., k_{liq} = const.  \text{ for} \quad \xi<0,\\
     \temperature_{gas}+\xi(\temperature_{gas})_\xi \quad & \text{and } \conductivity_{gas} = const., k_{gas} = const.  \text{ for} \quad \xi \geq 0,
  \end{cases}
\end{align}
we can use the solution of the dGRP from the previous section by setting
\begin{align}
  a_{liq/gas} = k_{liq/gas} \text{  and  } b_{liq/gas} = \conductivity_{liq/gas}.
\end{align}
Therefore, we assume a continuous temperature profile and a continuous energy flux and thereby enforce energy conservation. The numerical flux is finally given
by
\begin{align}
\fvec^*_{therm} &=\frac{2 \lambda^+ \lambda^- (\temperature^+-\temperature^-)}{\sqrt{\pi \Delta t}(\lambda^+\sqrt{k^-}+\lambda^-\sqrt{k^+})} + \frac{ \lambda^+
   \lambda^- (\sqrt{k^+}
\temperature^+_{\xi} + \sqrt{k^-} \temperature^-_{\xi})}{(\lambda^+\sqrt{k^-}+\lambda^-\sqrt{k^+})},
\label{eq:finheatflux}
\end{align}
with the shorthand notation $(*)^+=(*)_{gas}$ and $(*)^-=(*)_{liq}$.

\end{subsection}

\begin{subsection}{The Numerical Viscous Flux}
  \label{sec::viscoussub}
The second parabolic flux in the NSE refers to friction. Given the nature of the viscous stress tensor, this problem can not be reduced to one dimension and a
straightforward application of the dGRP, as was done for the numerical heat flux, is not possible. Therefore, we follow Gassner et al. \cite{Gassner2007a,Gassner2007},
in their approach for parabolic systems and extend it to the general liquid/vapor case considered here.

We investigate the viscous Riemann problem normal to the interface defined by the initial conditions \eqref{eq:grp} and the  partial differential equation
\begin{align}
   \pderivative{\uvec}{\zeit} - \pderivative{}{\xi_1} \left(\sum\limits_{j=1}^{3} \underline{\underline{D}}_{1j} \pderivative{}{\xi_j} \uvec \right) &= 0,
   \label{eq::viscsubb}
\end{align}
with the diffusion matrices,
 \begin{equation*}
       \underline{\underline{D}}_{11}=\frac{\viscosity}{\dens}
        \begin{pmatrix}
                                     0 &              0 &   0 &   0& 0\\
                       -\frac{4}{3}\vel{1} &    \frac{4}{3} &   0 &   0& 0\\
                                  -\vel{2} &              0 &   1 &   0& 0\\
                                  -\vel{3} &              0 &   0 &   1& 0\\
      -\frac{4}{3}(\vel{1})^2 -(\vel{2})^2 - (\vel{3})^2  & \frac{4}{3}\vel{1} & \vel{2} & \vel{3}& 0
        \end{pmatrix},
\end{equation*}
 \begin{equation*}
        \underline{\underline{D}}_{12}=\frac{\viscosity}{\dens}
        \begin{pmatrix}
                         0 &    0 &               0 & 0& 0\\
            \frac{2}{3}\vel{2} &    0 &    -\frac{2}{3} & 0& 0\\
                      -\vel{1} &    1 &               0 & 0& 0\\
                         0 &    0 &               0 & 0& 0\\
        -\frac{1}{3}\vel{1}\vel{2} &  \vel{2} & -\frac{2}{3}\vel{1} & 0& 0\\
        \end{pmatrix},
\end{equation*}
 \begin{equation*}
        \underline{\underline{D}}_{13}=\frac{\viscosity}{\dens}
        \begin{pmatrix}
                         0 &    0 & 0 &               0 & 0 \\
            \frac{2}{3}\vel{3} &    0 & 0 &    -\frac{2}{3} & 0 \\
                         0 &    0 & 0 &               0 & 0 \\
                      -\vel{1} &    1 & 0 &               0 & 0 \\
        -\frac{1}{3}\vel{1}\vel{3} &  \vel{3} & 0 & -\frac{2}{3}\vel{1} & 0
     \end{pmatrix}.
\end{equation*}
Here, $\xi_1$ denotes the coordinate into normal direction.  Note that the use of the diffusion matrices is equivalent to the standard flux form of the NSE
\cite{Gassner2007}.

From here, we further decompose the viscous flux into
\begin{align}
   \fvec_{visc}^*=\fvec_{visc,1}^*+\fvec_{visc,2}^*+\fvec_{visc,3}^*,
    \label{eq::visnumflux}
\end{align}
where $\fvec_{visc,1}^*=\fvec_{visc,1}^*(\uvec_{\xi_1}^{liq},\uvec_{\xi_1}^{gas})$ refers to the flux contribution of the normal gradients and
$\fvec_{visc,2}^*=\fvec_{visc,2}^*(\uvec_{\xi_2}^{liq},\uvec_{\xi_2}^{gas})$, $\fvec_{visc,3}^*=\fvec_{visc,3}^*(\uvec_{\xi_3}^{liq},\uvec_{\xi_3}^{gas})$ the
contribution of the tangential gradients. Assuming a continuous tangential gradient contribution, as in \cite{Gassner2007}, leads to an approximation of the tangential contributions as the arithmetic mean of the left and right fluxes
\begin{align}
     \fvec_{visc,2}^*&=\frac{1}{2}\left(\underline{\underline{D}}_{12,liq} \pderivative{}{\xi_2}\uvec_{liq} +\underline{\underline{D}}_{12,gas}
     \pderivative{}{\xi_2}\uvec_{gas}\right), \\
     \fvec_{visc,3}^*&=\frac{1}{2}\left(\underline{\underline{D}}_{13,liq} \pderivative{}{\xi_3}\uvec_{liq} +\underline{\underline{D}}_{13,gas}
     \pderivative{}{\xi_3}\uvec_{gas}\right).
      \label{eq::tangential_viscous}
\end{align}
Note that this assumption is in line with the integral jump conditions across a non-evaporating interface under the neglection of Marangoni forces.

For the normal contribution we can use a more sophisticated approach based on the approach of \citet{Gassner2007}. We linearize the diffusion matrix of the normal
contribution $\underline{\underline{D}}_{11}$ around a mean state $\uvec_{m}$ leading to a linear equation  system:
\begin{align}
   \pderivative{\uvec}{\zeit} -\pderivative{}{\xi_1}  \left(\underline{\underline{D}}_{11}(\uvec_{m}) \pderivative{}{\xi_1} \uvec\right)  &= 0.  \label{eq:linvisc}
\end{align}
The linearized diffusion matrix can be diagonalized to the matrix of eigenvalues
\begin{align}
 \underline{\underline{\boldsymbol\Lambda}} =
 \begin{pmatrix}
   0 & 0 & 0 & 0 & 0\\
   0 & \frac{4}{3}\frac{\viscosity}{\dens^m} & 0 & 0 & 0\\
   0 & 0 & \frac{\viscosity}{\dens^m} & 0 & 0\\
   0 & 0 & 0 & \frac{\viscosity}{\dens^m} & 0\\
   0 & 0 & 0 & 0 & 0
 \end{pmatrix},
 \end{align}
 with the corresponding matrix of eigenvectors
 \begin{align}
       \underline{\underline{\bf T}}=
          \begin{pmatrix}
                                                1  &                                       0 &                                      0 &                                       0 & 0\\
           \frac{m^m_1}{\dens^m} &                                       1 &                                      0 &                                       0 & 0\\
           \frac{m^m_2}{\dens^m} &                                       0 &                                      1 &                                       0 & 0\\
           \frac{m^m_3}{\dens^m} &                                       0 &                                      0 &                                       1 & 0\\
           \frac{E^m  }{\dens^m} & \frac{m^m_1}{\dens^m} & \frac{m^m_{2}}{\dens^m}& \frac{m^m_3}{\dens^m} & 1\\
        \end{pmatrix},
       \label{eq:eigv_thermal}
 \end{align}
 where the superscript $m$ refers to the mean values. The characteristic form of  Eq. \eqref{eq:linvisc} then follows as
\begin{align}
    \pderivative{\wvec}{\zeit} - \pderivative{}{\xi_1} \left(\underline{\underline{\boldsymbol\Lambda}} \pderivative{}{\xi_1} \wvec\right)  &= 0
    \label{eq:linvisc2}
 \end{align}
 with the vector of characteristic variables and the vector of their derivatives
 \begin{align}
  \wvec = \underline{\underline{\bf T}}^{-1} \uvec \text{ and  } \pderivative{}{\xi_1} \wvec = \underline{\underline{\bf T}}^{-1} \pderivative{}{\xi_1} \uvec.
\end{align}
By applying this transformation procedure on both sides of the phase boundary, we obtain a decoupled systems of diffusion problems with discontinuous
coefficients normal to the interface. The corresponding generalized diffusive Riemann problem has the initial data
\begin{align}
   \wvec(\xi_1,0)=
   \begin{cases}
      \wvec_{liq}+\xi_1 (\wvec_{\xi_1})_{liq} \quad  \text{ for} \quad \xi_1<0\\ \wvec_{gas}+\xi_1  (\wvec_{\xi_1})_{gas} \quad  \text{ for} \quad \xi_1\geq0
   \end{cases}.
\end{align}

Note that the same transformation matrix is used on both sides of the phase boundary, while the eigenvalue matrix differs and thus introduces discontinuous
coefficients. Now, we can use Eq. \eqref{eq:intheatflux} to obtain the components $J_i$ of the numerical viscous flux by choosing
\begin{align}
  a_{liq/gas} = \Lambda_{ii,liq/gas} \text{  and  } b_{liq/gas} = \Lambda_{ii,liq/gas} \text{  for  } i =1\dots,5.
\end{align}
They are given by
\begin{equation}
J_i=\frac{2 \Lambda_{ii}^+ \Lambda_{ii}^- (W_i^+- W_i^-)}{\sqrt{\pi \Delta t}(\Lambda_{ii}^+\sqrt{\Lambda_{ii}^-}+\Lambda_{ii}^-\sqrt{\Lambda_{ii}^+})}
+ \frac{ \Lambda_{ii}^+ \Lambda_{ii}^- (\sqrt{\Lambda_{ii}^+} W^+_{\xi_1} + \sqrt{\Lambda_{ii}^-}
W^-_{\xi_1})}{(\Lambda_{ii}^+\sqrt{\Lambda_{ii}^-}+\Lambda_{ii}^-\sqrt{\Lambda_{ii}^+})},
   \label{eq:jdgrp}
\end{equation}
where we again use the shorter notation of $(*)^+=(*)_{gas}$ and $(*)^-=(*)_{liq}$.  The resulting fluxes are transformed back to the conservative variables by
multiplying with the transformation matrix
\begin{equation}
   \fvec_{visc,1}^*=\underline{\underline{\bf T}}\mathbf{J}.
   \label{eq:fvisc1}
\end{equation}
The total, numerical viscous flux is then given by \cref{eq::visnumflux}.

So far we have not discussed the choice of $\uvec_m$ for the linearization. For small jumps in the coefficients it can be chosen simply as the arithmetic mean
of the liquid and vapor states as in \cite{Gassner2007}. For larger jumps, as is considered in this work, an average state incorporating the different dynamic
viscosities is more favorable. We follow the structure of the exact solution at $\xi=0$ of a scalar dGRP with jumping coefficients and choose
\begin{equation}
  \uvec_{m}=\dfrac{ \sqrt{\viscosity_{liq}} \uvec_{liq}+\sqrt{\viscosity_{gas}}\uvec_{gas}}{\sqrt{\viscosity_{liq}} + \sqrt{\viscosity_{gas}}}.
   \label{eq:wmean}
\end{equation}
\end{subsection}

\end{section}
\begin{section}{Results}
\label{sec:results}
In the present section, numerical results are presented which validate the proposed interfacial Riemann solvers and display their capabilities. We begin
with a discussion on the thermal Riemann solver, followed by the viscous solver.

\subsection{Thermal Riemann Solver}
In the present section, the interfacial heat conduction are validated. Therefore, we employ the well-known test case of Sod in a modified form with two heat
conducting fluids. The initial conditions are given by
\begin{equation}
   \uvec(\xi,0)=
     \begin{cases}
        (1,0,0,0,1)^{\mathrm{T}} \, &\text{for} \quad \xi<0, \\
        (0.125,0,0,0,0.1)^{\mathrm{T}} \, &\text{for} \quad \xi\ge0.
     \end{cases}
   \label{eq:sod}
\end{equation}
At $\xi=0$, we position an interface, which separates the two fluids from each other. Each fluid obeys a perfect gas EOS with the specific heat capacity ratio of
$\gamma=1.4$ and a specific heat capacity at constant volume of $c_v=1$. Although obeying the same equation of state, the two fluids differ in their thermal
conductivity with $\lambda=0.1$ for the
fluid left of the interface and $\lambda=0.001$ for the fluid right of the interface.

No analytical solution is available for this problem, therefore we compare the sharp-interface method with a single-phase DGSEM Navier-Stokes solver
\cite{Krais2020}. This
method is able to consider variable thermal conductivities through the BR1 lifting procedure \cite{Bassi1997} and is therefore a  good reference to validate the
novel ghost fluid approach. The solution was obtained with $N_{\mathrm{Elems}}=200$ solution elements and a polynomial degree of $N=3$, which was sufficient to
achieve grid convergence.  We compare
this reference solution  with the
discussed sharp-interface method using the thermal interfacial Riemann solver and the HLLC solver. To focus on the performance of the Riemann solvers we further
employ a moving mesh approach for this one-dimensional simulation as was done in \cite{Hitz2020,joens2023,muller2023}. In both simulations we neglect viscous effects.

Four different meshes $N_{\mathrm{Elems}}=25$, $N_{\mathrm{Elems}}=50$, $N_{\mathrm{Elems}}=100$ and $N_{\mathrm{Elems}}=200$  were used as well as a polynomial
degree of $N=3$. The results are depicted in Fig. \ref{fig:heatft} at $t=0.2$.  The solution appears to be grid converged with $N_{\mathrm{Elems}}=100$. The converged
solution of the sharp-interface method shows a perfect agreement with the reference data, implying an accurate coupling of the advection and thermal conduction
across the interface.

\begin{figure}
   \centering
   \includegraphics[width=\linewidth]{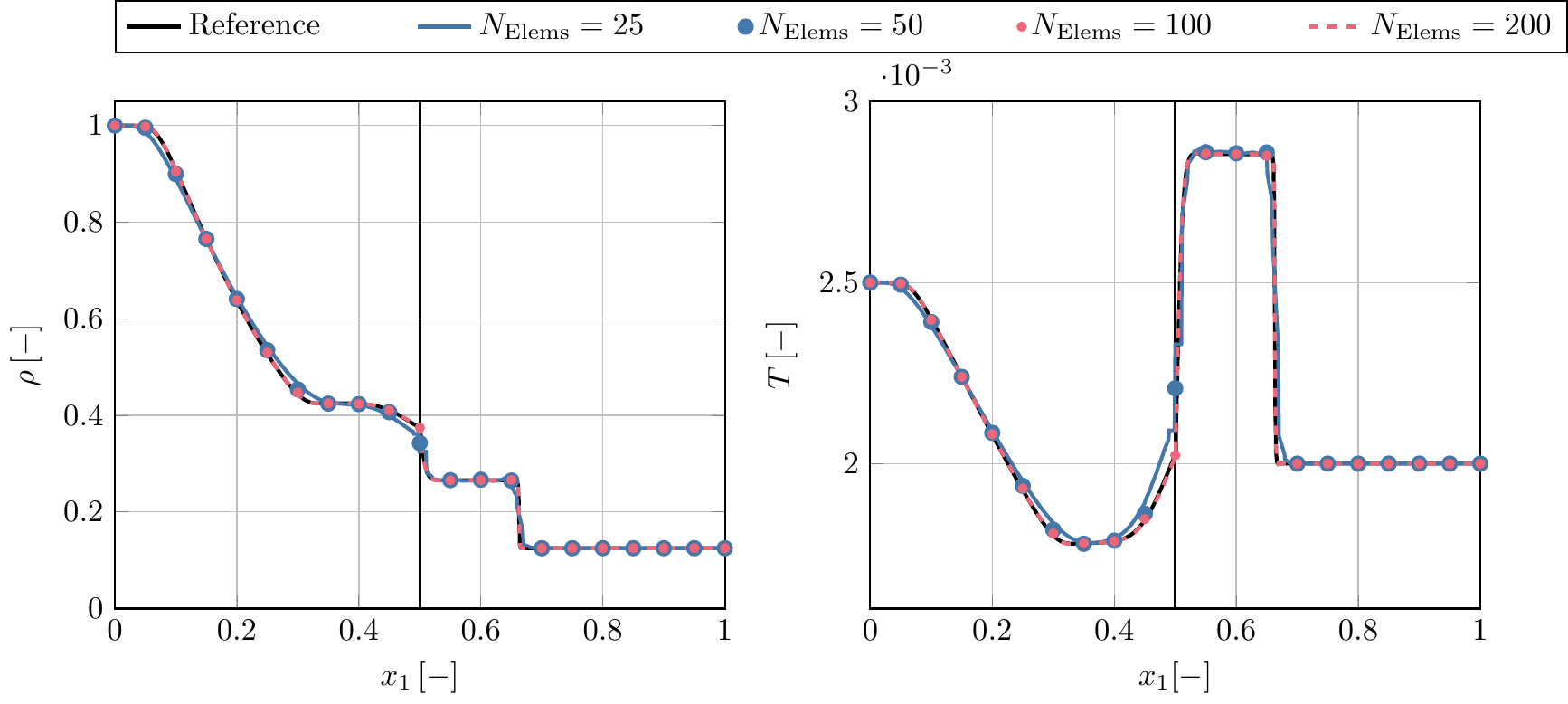}
   \caption{Sod problem with heat conduction after $ t =0.2$. Solutions of the sharp-interface method with the thermal Riemann solver and different meshes
      are shown compared to a numerical reference solution. The interface is indicated in black.}
   \label{fig:heatft}
\end{figure}

As second test case, we consider a water droplet in air impinging on a water surface, similar to the case
in \cite{zeifang2021low}. Here, we focus on the effects of thermal conduction. Droplet, air and water surface are  initialized with
a different temperature, with the droplet having the hottest temperature and the air the coldest one. A  summary of the geometrical setup, is shown in Fig.
\ref{fig:coffee}. Gravitation is introduced as an external field and is included as a source term.
The water is modeled with a stiffened gas EOS, using $\gamma=2.51$, a specific heat capacity ratio at constant pressure of
$c_p=4267\si{\joule\per\kilogram\per\kelvin}$, the stiffness parameter  $p_{\infty}=10^9 \si{\pascal}$ and a thermal conductivity of
$\lambda=0.6\si{\watt\per\meter\per\kelvin}$. The air is
modeled as a perfect gas with $\gamma=1.4$,  $c_p=1000\si{\joule\per\kilogram\per\kelvin}$ and $\lambda=0.024\si{\watt\per\meter\per\kelvin}$. The surface tension coefficient was set to $\sigma=58\cdot10^{-3}\si{\newton\per\meter}$.

\begin{figure}
\includegraphics{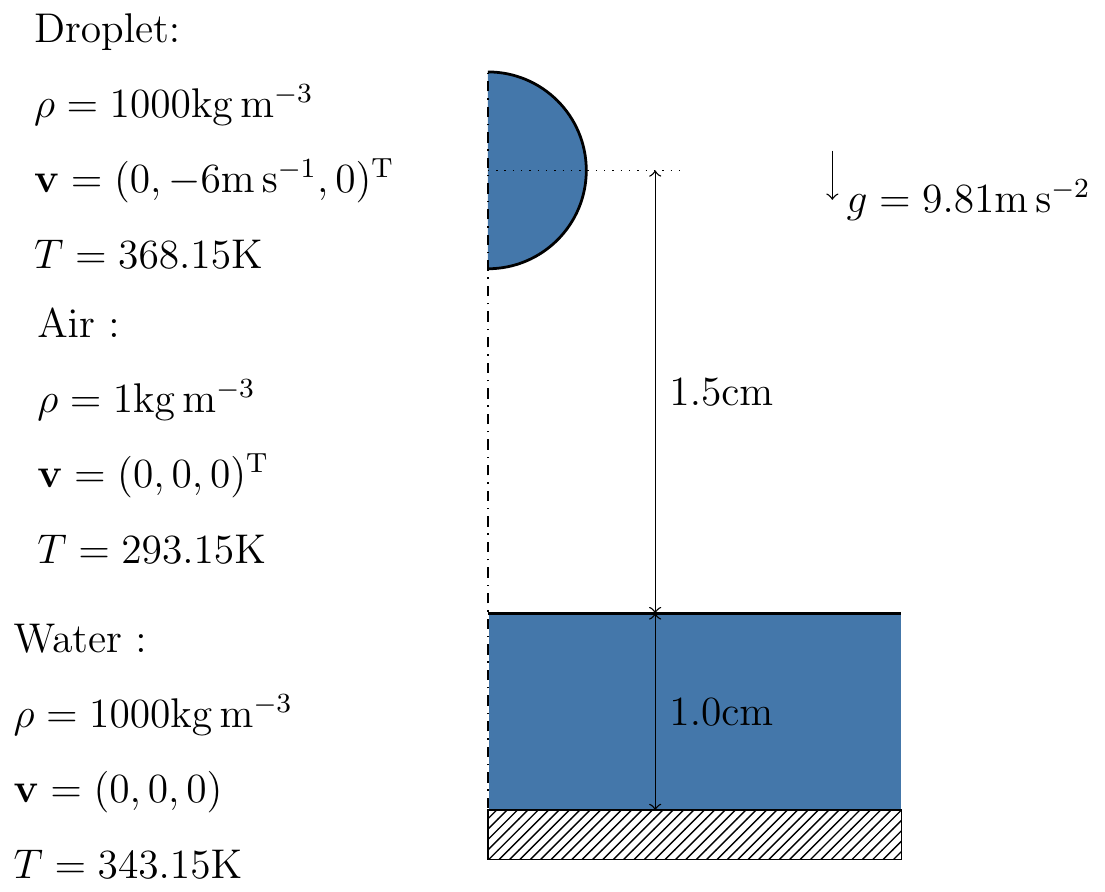}
   \caption{Setup of the impinging droplet test case.}
   \label{fig:coffee}
\end{figure}

\begin{figure}
\includegraphics{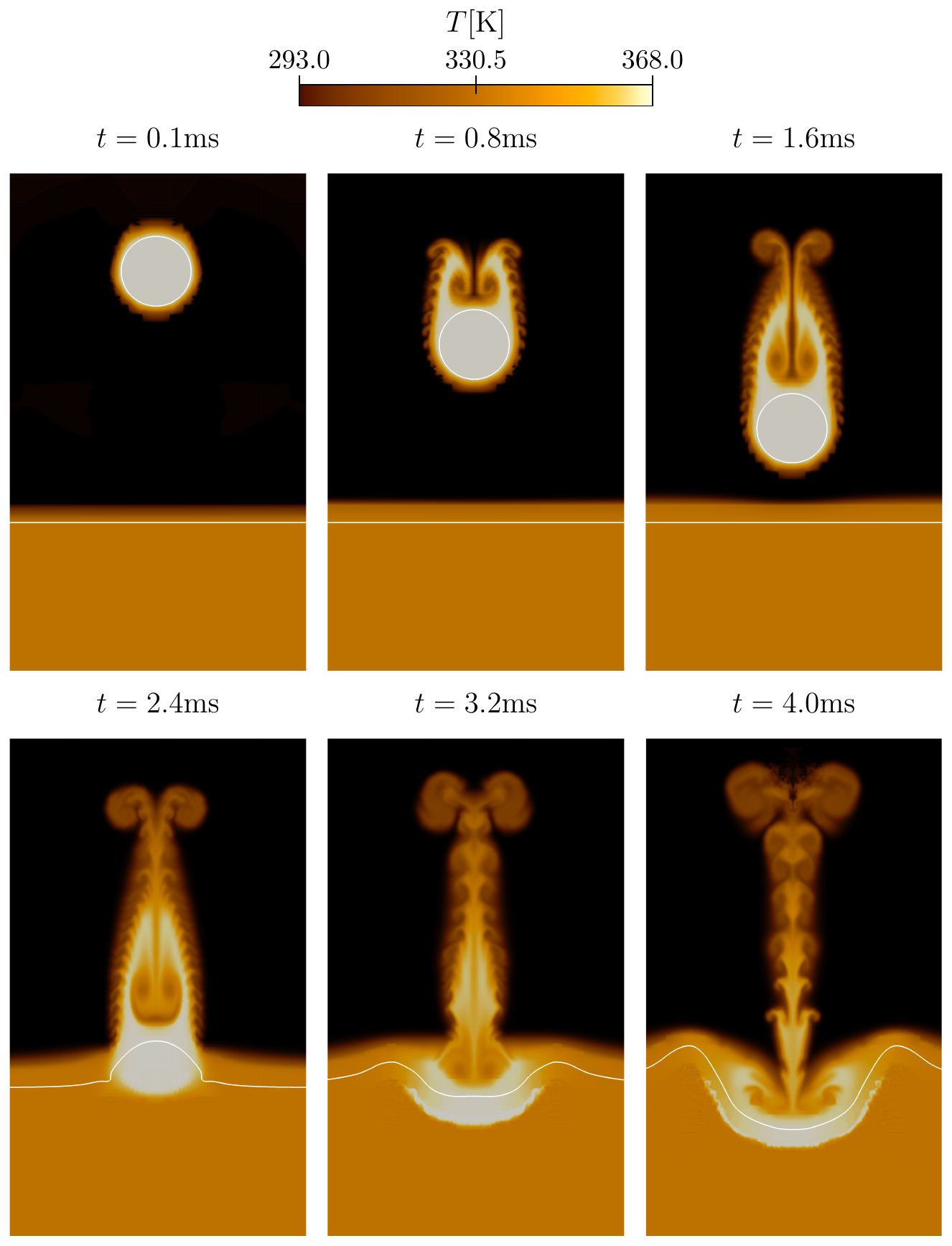}

   \caption{Temperature distribution of the impinging droplet at different time instances. Interfaces are given as a white line.}
   \label{fig:coffetemp}
\end{figure}

The two-dimensional domain  $\Omega\in[0,3\si{\centi\m}]^2$  was discretized with $120$ elements in $x_1$ direction and $60$ elements in $x_2$ direction. A
polynomial degree of $N=5$ was used. The Lax-Friedrichs Riemann solver was used in the bulk phases. At the interface, the HLLC solver in conjunction with the
novel thermal Riemann solver was employed. Viscous effects were neglected. As boundary conditions, an Euler wall at the lower boundary was employed while the
remaining boundaries used a Dirichlet condition based on the initial data.

The temperature distribution for different time instances is shown in Fig. \ref{fig:coffetemp}.  After $t=0.1\si{\milli\second}$, the droplet has slightly moved
downwards. Heat transfer across the interface induces a heating of the surrounding air. Similar observation can be made at the liquid surface. After
$t=0.8\si{\milli\second}$, a thermal boundary layer at the surface has developed. The droplet moved further downwards and has left a trailing wake of hot air.
Therein, buoyancy driven instabilities formed in the areas where hot and cold air meet. After $t=1.6 \si{\milli\second}$, the droplet is very close to the
liquid surface. The thermal boundary layer of the liquid surface has been slightly quenched by the incoming
pressure waves from the  arriving droplet. At $t=2.4 \si{\milli\second}$, the droplet partially merged into the surface, raising the thermal boundary layer.
After $t=3.2\si{\milli\second}$ the droplet is fully engulfed in the liquid surface. Below, the hot liquid starts to spread. The wake above the surface has
further cooled and an intricate buoyancy driven flow pattern has emerged. Finally, at $t=4\si{\milli\second}$, a narrowly defined area of hot air remains in the
wake region, in which vortices dominate the flow field. At the droplet's initial position, the onset of a Rayleigh-Taylor instability is visible.  Waves on the
liquid surface have further traveled away from the center line and the  hot liquid from the droplet has now spread almost below the entire visible surface.

The depicted temperature distribution is mainly driven by the interfacial exchange of heat across the interface.  The temperature in the air would
have been nearly uniform in a simulation without heat conduction. Due to the heating of the air by the droplet, a buoyancy driven flow was induced. The
observations made above are in line with the physical expectations and therefore demonstrate the capability of the proposed method.

\subsection{Viscous Riemann Solver}
In the this section, the viscous Riemann solver is considered. First, we solve a quasi-one-dimensional problem: a multi-material adaption of the
first problem of Stokes. A discussion of the original problem can be found in \cite{Schlichting2017a,Dumbser2018}. We choose the initial conditions as follows:
\begin{equation}
   (\rho,  v_1,  v_2, v_3, p ) =
   \begin{cases}
      (1, 0,-0.1,0,1) \, & \text{for} \quad x_1<0, \\
      (1, 0, 0.1,0,1) \, & \text{for} \quad x_1>0.
   \end{cases}
   \label{eq:stokes}
\end{equation}
A different fluid is present at each side of the discontinuity positioned at  $x_1=0$. As done before, these fluids are described by the same EOS, a perfect
gas EOS with $\gamma=1.4$ and
$ c_v =1$. However, they  differ in their dynamic viscosity coefficient, given by
\begin{equation}
    \mu =
   \begin{cases}
      0.01 \, & \text{for} \quad x_1<0, \\
      0.001 \,& \text{for} \quad x_1>0.
   \end{cases}
   \label{eq:stokesvisc}
\end{equation}
Similarly to the original first problem of Stokes, the Navier-Stokes-equations can be reduced to a one-dimensional equation due to the prescribed initial data,
reading
\begin{equation}
   \frac{\partial v_2}{\partial t}= \mu \frac{\partial^2 v_2}{\partial {x}_1^2}.
   \label{eq:NSstokes}
\end{equation}
Eq. \eqref{eq:NSstokes} is of the form of Eq. \eqref{eq::diff}. Therefore, the solution of this problem is given by the solution of the dGRP as described in
section \ref{sec::dgrp_diff}.

The modified first problem of Stokes is considered with the present sharp-interface method. The employed two-dimensional computational domain was $[0,1]\times[0,1]$, where non-reflecting outflow boundaries were employed at the $x_1$ borders, while periodic boundary conditions were used at the
$x_2$ boundaries. We chose a polynomial degree of $N=3$,  the HLLC solver was applied to calculate the hyperbolic interface flux and the viscous Riemann solver
for the parabolic
flux. Heat conduction effects were neglected in this case.

\begin{figure}
   \centering
   \includegraphics[width=.7\linewidth]{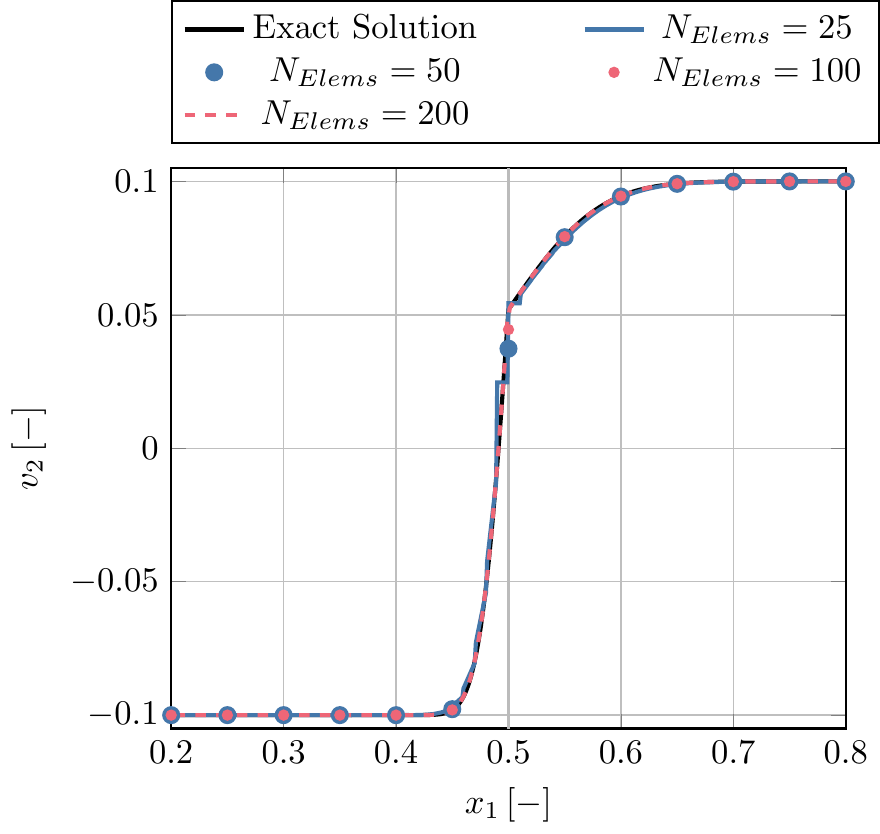}
   \caption{Solution of the modified first problem of Stokes. Comparison of the sharp interface method with the exact solution.}
   \label{fig:Stokes}
\end{figure}

The solution of the sharp-interface method as well as the exact solution is shown in Fig. \ref{fig:Stokes} for four different meshes:  $N_{Elems}=25$,
$N_{Elems}=50$, $N_{Elems}=100$ and $N_{Elems}=200$ elements. The sharp-interface method shows grid convergence for $N_{Elems}=100$. Additionally, the numerical
results are in perfect agreement with the exact solution.

As a more complex test case, a three-dimensional droplet oscillation is investigated as a second test case. Taken from the work of \citet{Lalanne2015a}, the
initial contour of the droplet is given by
\begin{equation*}
   \psi(\vec{x})=\sqrt{(x_1^2+x_2^2+x_3^2)} - R_D \left[1+ 0.15\left(3\cos^2 \left(\frac{x_1}{\sqrt{x_1^2+x_2^2+x_3^2}}\right)-1\right)\right],
   \label{eq:drop}
\end{equation*}
with the equilibrium droplet radius $R_D$. The oscillation period of this elongated droplet in air has been studied by \citet{Prosperetti1980} and
\citet{lamb1924hydrodynamics} and is given by
\begin{equation}
   T_{osc} = \frac{2\pi \sqrt{R_D^3 (2n \rho_{air} + (n+1)\rho_{liq})}}{\sqrt{\sigma  (n-1)n(n+1)(n+2)}},
   \label{eq:oscperiod}
\end{equation}
where $n$ refers to the oscillation mode, with $n=2$ being the dominating mode. The oscillation of the droplet is damped over time by the effects of viscosity.
The decay of the amplitude thereby reads as
\begin{equation}
   \frac{r_D(t)}{r_{D,i}(t)} =\exp \left( -t \frac{\mu_{liq}}{\rho_{liq}}\frac{(n-1)(2n+1)}{R_D^2}\right),
   \label{eq:oscdamp}
\end{equation}
where $r_D(t)$ denotes the time dependent droplet radius and $r_{D,i}(t)$ the theoretical droplet radius in the inviscid case.

In the following, the sharp-interface method is used to simulate such a viscous, oscillating droplet in non-dimensional units. The droplet was positioned inside the center of the
computational domain $\Omega\in[-2.5,2.5]^3$ with a density of $\rho_{liq}=1000$, a pressure of $P_{liq}=500$ and a dynamic viscosity of
$\mu_{liq}=10$. The surrounding air was initialized with a density of $
\rho_{air}=1$, a pressure of $P_{air}=1$ and a dynamic viscosity of $\mu_{air}=0.01$. The liquid was modeled by a stiffened gas EOS with $\gamma=4$ and the
stiffness parameter
$P_{\infty} =10^3$. The air was described with a perfect gas EOS using $\gamma=1.4$. A surface tension coefficient for the liquid-air interface of $\sigma=250$
was used. The sharp-interface method used a polynomial degree of $N=3$ as well as the HLLC solver for the hyperbolic interfacial Riemann problem and the proposed viscous
Riemann solver for the parabolic part. Heat conduction was neglected and periodic boundary conditions were used.

The temporal variation of the droplet radius on the $x_1$-axis is depicted in Fig. \ref{fig:oscdrop} for three different mesh sizes $N_{Elems}=10^3$,
$N_{Elems}=20^3$ and $N_{Elems}=30^3$ and the theoretically predicted
envelope of the oscillation frequency from Eq. \eqref{eq:oscdamp}. The results clearly show a convergence of the scheme towards the theoretical
prediction, both in the oscillation frequency as well as the damping rate. Over time, the
oscillation frequency reduces and the damping increases. This an effect of an increased numerical dissipation induced by the bulk phase Riemann solvers for the present
low-speed problem, as was discussed by \citet{zeifang2021low}. Therefore, the precise prediction of the damping rate in the first oscillation period is a strong indication for the validity of the proposed method.

\begin{figure}
   \centering
   \includegraphics[width=.7\linewidth]{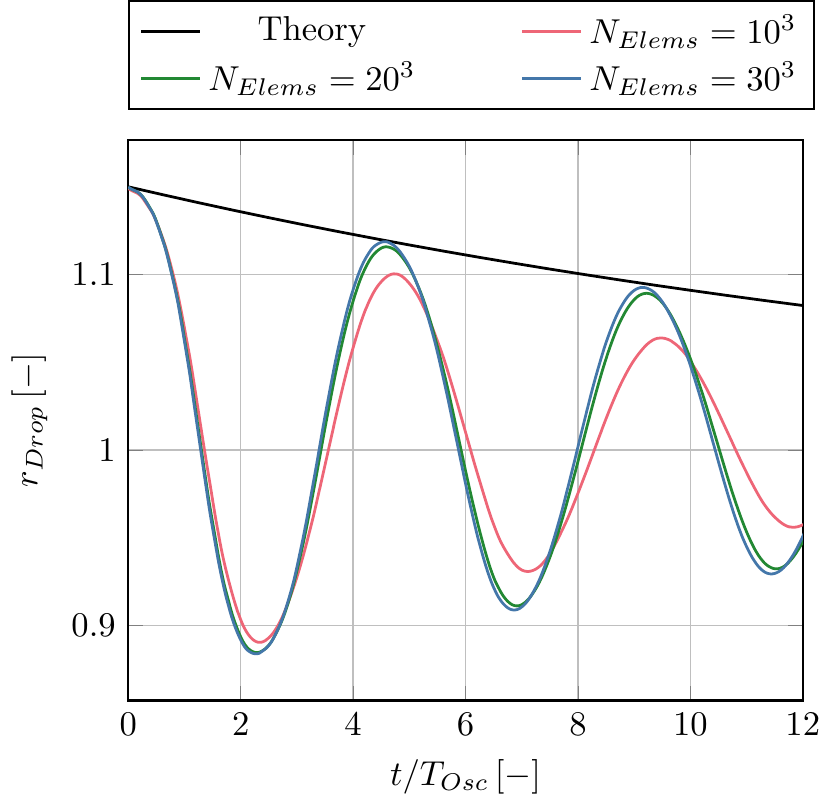}
   \caption{Droplet radius oscillation on the $x_1$-axis over several oscillation periods supplemented by the envelope of the analytically predicted damped
   radius. }
   \label{fig:oscdrop}
\end{figure}

Finally, we simulated a 2D shock-droplet interaction in the shear-induced entrainment (SIE) regime with Weber number $\text{We} = 7339$ inspired by
\citet{winter2019numerical}.
Initially, a droplet with the diameter of $D_0=1\text{m}$ is positioned at $(x,y) = (0,0)$ inside the computational domain $\Omega = [-2D_0,10D_0] \times
[0,3D_0]$, with $D_0$ being the droplet radius. The surrounding air is at rest. A right-moving shock wave,
with Mach number $\text{Ma}=1.47$, is placed at $x=-D_0$. Hence, at the left boundary a Dirichlet boundary condition is imposed based on the post shock state. The
lower boundary is set as a symmetry plane and the remaining boundaries are treated with non-reflective outflow conditions. A summary of the initial conditions and material
parameters is given in Table \ref{Tab:ini_cond_sie}. The initial pressure difference between the droplet and air was defined by the Young-Laplace
Law
\begin{equation}
\Delta p  = \dfrac{2 \sigma}{D_0}.
	\label{eq:youngLaplace}
\end{equation}
The dynamic viscosities of water and air were calculated by using the Ohnesorge number
\begin{equation}
   \text{Oh} = \dfrac{\mu_{liq}}{\sqrt{\rho_{liq}D_0\sigma}},
\label{eq:ohnesorge}
\end{equation}
the Weber number
\begin{equation}
   \text{We} = \dfrac{\rho_{gas}u_{gas}^2D_0}{\sigma},
\label{eq:weber}
\end{equation}
and the Reynolds number
\begin{equation}
   \text{Re}  = \dfrac{\rho_{gas} u_{gas} D_0}{\mu_{gas}}.
\label{eq:reynolds}
\end{equation}
Following in \citet{winter2019numerical}, the Ohnesorge number is taken as $\text{Oh}=6.9\times 10^{-3}$ and the Reynolds number as
 $\text{Re}=229485$.

\begin{table}[h]
\centering
  \begin{tabular}{ c  c  c  c  c  c  c}
    \hline
    	Fluid & $p_0 [\si{\pascal}]$ & $\rho_0 [\si{\kilogram\per\cubic\meter}]$ & $\gamma[-]$ & $p_\infty [\si{\giga\pascal}]$ & $\mu [10^{-2} \si{\pascal\second}]$ & $\sigma[10^{-2}\si{\kilo\gram\per\square\second}]$ \\ \hline
    Air & 101325 & 1.204 & 1.4 & 0 & 0.011\\
    Water & 101325.146 & 1000 & 6.12 & 0.343 & 5.885 & 7.28\\
    \hline
  \end{tabular}
\caption{Initial conditions and material parameters for the shock-droplet interaction}
\label{Tab:ini_cond_sie}
\end{table}

The domain  was discretized with $512 \times 256$  elements. The polynomial degree of the DG solution was $\textit{N}=3$. We used the HLLC Riemann solver in the
bulk and an explicit 4th order Runge-Kutta time stepping scheme.  At the interface, the HLLC solver and the viscous Riemann solver was employed. Heat conduction
was neglected in the entire domain.  A time series of the obtained results is given in Figure \ref{fig:sie}. The solution is plotted at the
non-dimensional time
\begin{equation}
t^* = \dfrac{t}{\dfrac{D_0}{u_s}\sqrt{\dfrac{\rho_{liq}}{\rho_s}}}
\label{eq:nonDimTime}
\end{equation}
with $u_s$ and $\rho_s$ denoting the post-shock velocity and density of air. Figure \ref{fig:sie} shows the numerical Schlieren and non-dimensional streamwise
velocity $u^*=u/u_s$ at $t^*=0.25$, $t^*=0.5$ and $t^*=1.0$. Due to the striking of the shock wave and the acceleration of the air, the droplet deforms.
Separation of the flow from the droplet surface results in the shedding of
vortices. Comparing the present results to the ones given by \citet{winter2019numerical}, the droplet deformation and generated vorticial structures agree very
well with each other.

\begin{figure}[]
\vspace{-3cm}
    \centering
    \begin{subfigure}[t]{1.0\textwidth}
    \centering
       \includestandalone[width=.65\textwidth,trim= 50 000 100 000, clip]{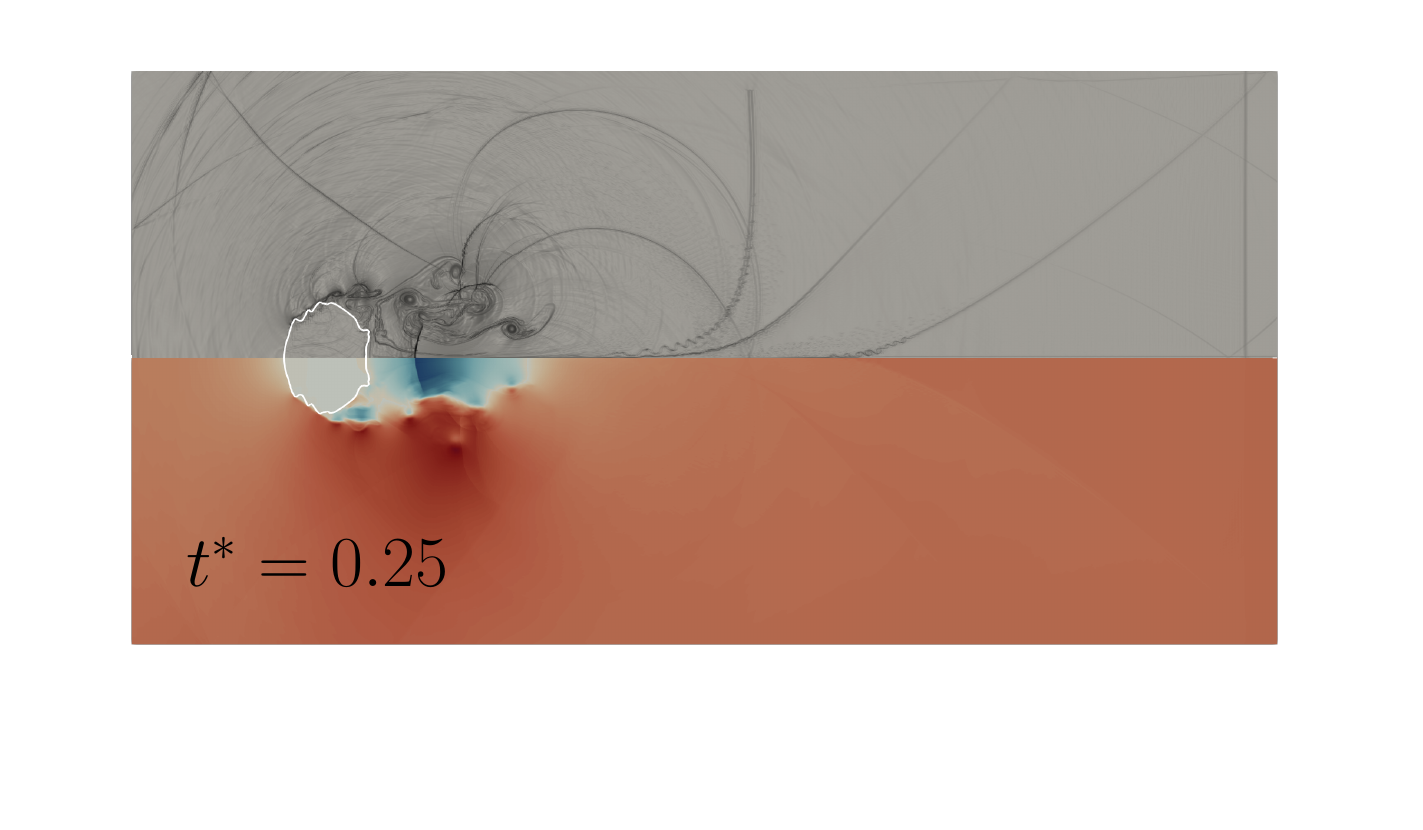}
    \end{subfigure}

    \vspace{-1.75cm}
    \begin{subfigure}[t]{1.0\textwidth}
    \centering
       \includestandalone[width=.65\textwidth,trim= 50 000 100 000, clip]{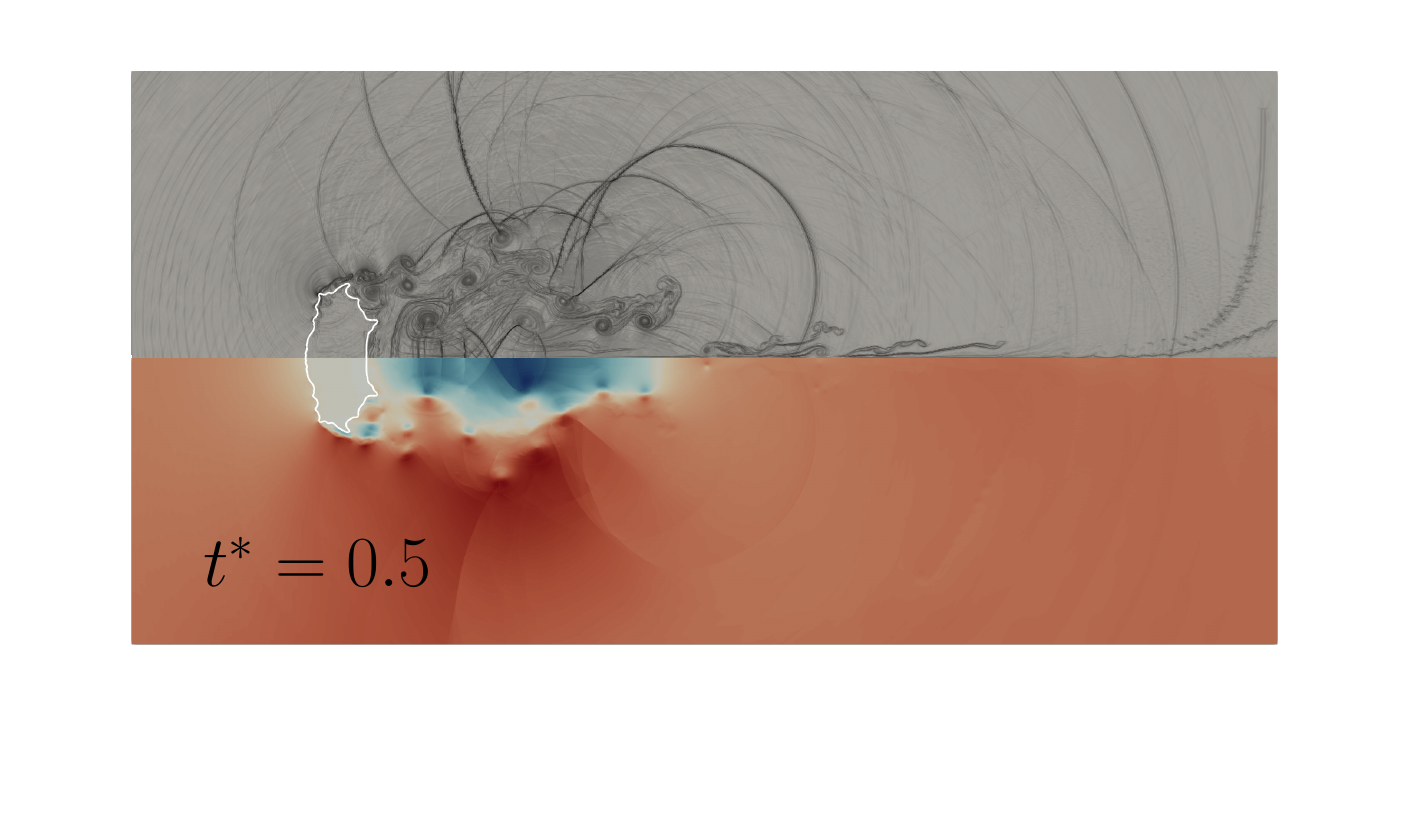}
    \end{subfigure}

    \vspace{-1.75cm}
    \begin{subfigure}[t]{1.0\textwidth}
    \centering
       \includestandalone[width=.65\textwidth,trim= 50 000 100 000,clip]{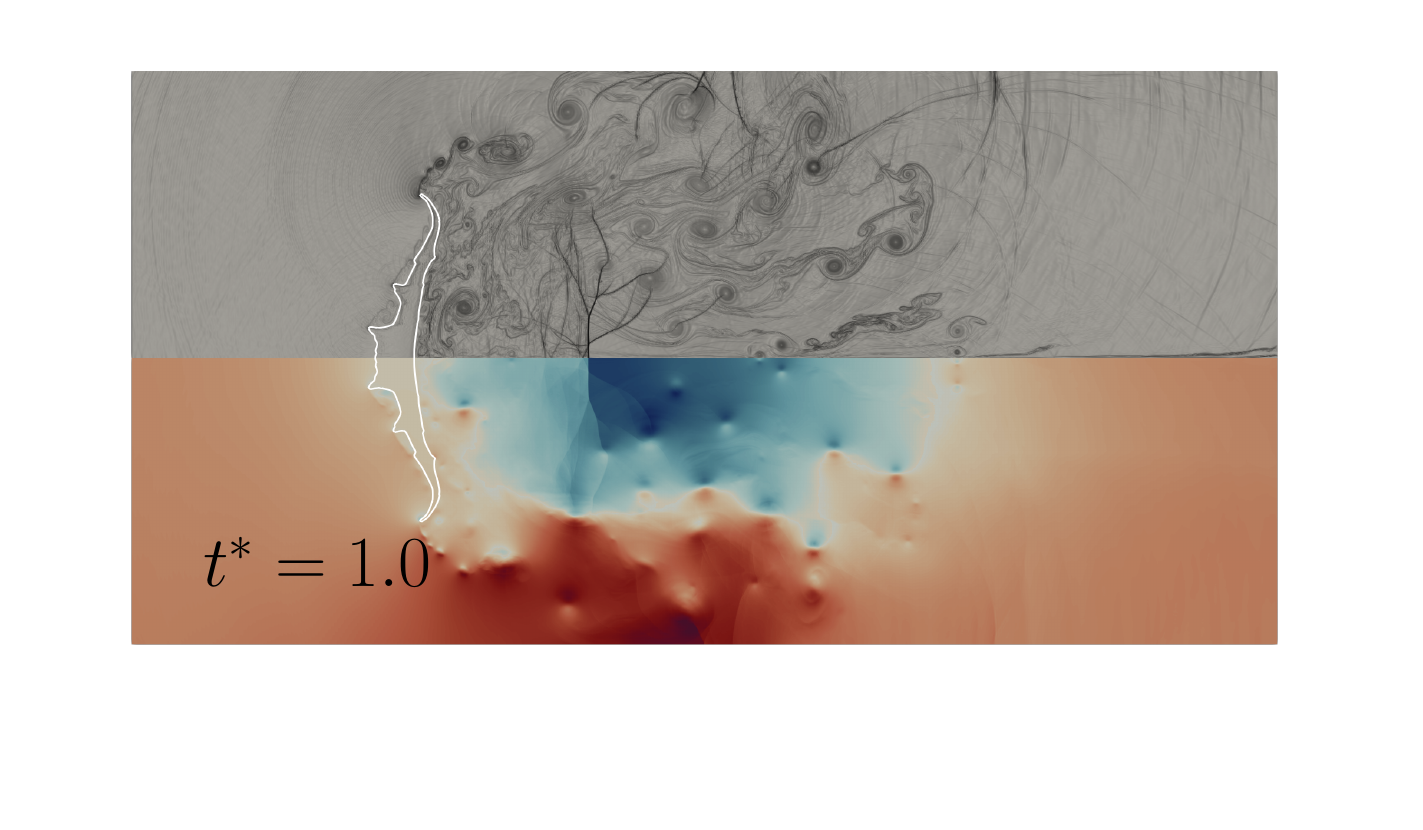}
    \end{subfigure}

    \vspace{-1.75cm}
    \includestandalone[width=0.45\textwidth]{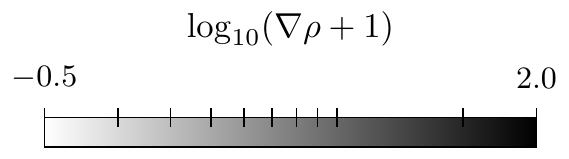}
    \includestandalone[width=0.45\textwidth]{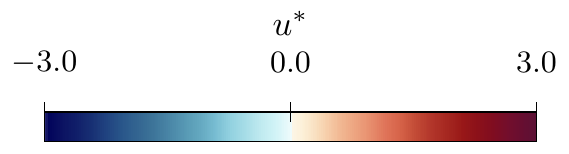}
    \caption{Shock-droplet interaction with numerical Schlieren at the top, non-dimensional streamwise velocity $u^*$ at the bottom. The interface is displayed as a white line.}
    \label{fig:sie}
  \end{figure}

\end{section}

\begin{section}{Conclusion}
\label{sec:conclusion}
In the present work we discussed a novel approach based on Riemann-solvers for the consideration of heat conduction and friction at sharp-interfaces. We extended
the ghost fluid method by defining the interfacial fluxes as the sum of the hyperbolic flux and the two parabolic fluxes, similar to the typical single-fluid
case. Each of the fluxes is individually defined by a Riemann solver. We employed the well-known HLLC solver for the hyperbolic flux and formulated novel
solvers based on the diffusive generalized Riemann problem. For the thermal Riemann problem, we were able to simplify the
problem to a scalar equation and thus derived an exact interfacial thermal flux. Such an approach is not possible for the viscous Riemann problem, for which we
derived an approximate solution based on a linearization of the diffusion matrix.

The novel Riemann solvers were employed in a compressible sharp-interface level-set ghost fluid method. The fluid flow in the bulk phases was solved with the
DGSEM with finite volume shock capturing. We studied one-dimensional problems dominated by heat conduction and friction and observed an excellent agreement of
our approach
with the employed reference data. Additionally, we investigated the viscous damping of an oscillating droplet, for which our scheme was able
to recover the theoretical predictions. The robustness and applicability of the numerical scheme was demonstrated in two separate cases: a heated falling droplet
impinging on a water surface and a shock-droplet interaction in the shear-induced entrainment regime.

Future investigations of the proposed method should focus on a comparing it with other well-known approaches. Further, other parabolic
effects, e.g. species diffusion, could be easily included by following the outline of the present work. Finally, we want to use the inclusion of viscous and
thermal effects in the study of droplet grouping effects.

\end{section}
\begin{section}{Acknowledgements}

\label{sec:acknowledgements}
The authors also kindly acknowledge the financial support provided by the German Research Foundation (DFG) through the Project SFB-TRR 75, Project number
84292822 - ``Droplet Dynamics under Extreme Ambient Conditions'', the GRK 2160/1 "Droplet Interaction Technologies" under the project number 270852890 as well
as Germany's Excellence Strategy - EXC 2075 – 390740016. The simulations were performed on the national supercomputer HPE APOLLO (HAWK) at the High Performance
Computing Center Stuttgart (HLRS) under the grant number hpcmphas/44084.
\end{section}
\bibliographystyle{elsarticle-num-names}
\bibliography{literature}

\begin{thebibliography}{38}
\expandafter\ifx\csname natexlab\endcsname\relax\def\natexlab#1{#1}\fi
\providecommand{\url}[1]{\texttt{#1}}
\providecommand{\href}[2]{#2}
\providecommand{\path}[1]{#1}
\providecommand{\DOIprefix}{doi:}
\providecommand{\ArXivprefix}{arXiv:}
\providecommand{\URLprefix}{URL: }
\providecommand{\Pubmedprefix}{pmid:}
\providecommand{\doi}[1]{\href{http://dx.doi.org/#1}{\path{#1}}}
\providecommand{\Pubmed}[1]{\href{pmid:#1}{\path{#1}}}
\providecommand{\bibinfo}[2]{#2}
\ifx\xfnm\relax \def\xfnm[#1]{\unskip,\space#1}\fi
%Type = Article
\bibitem[{Fedkiw et~al.(1999)Fedkiw, Aslam, Merriman, and Osher}]{Fedkiw1999}
\bibinfo{author}{R.~P. Fedkiw}, \bibinfo{author}{T.~Aslam},
  \bibinfo{author}{B.~Merriman}, \bibinfo{author}{S.~Osher},
\newblock \bibinfo{title}{A non-oscillatory {E}ulerian approach to interfaces
  in multimaterial flows (the ghost fluid method)},
\newblock \bibinfo{journal}{Journal of Computational Physics}
  \bibinfo{volume}{152} (\bibinfo{year}{1999}) \bibinfo{pages}{457--492}.
  \DOIprefix\doi{10.1006/jcph.1999.6236}.
%Type = Article
\bibitem[{Hirt and Nichols(1981)}]{Hirt1981}
\bibinfo{author}{C.~Hirt}, \bibinfo{author}{B.~Nichols},
\newblock \bibinfo{title}{Volume of fluid ({VOF}) method for the dynamics of
  free boundaries},
\newblock \bibinfo{journal}{Journal of Computational Physics}
  \bibinfo{volume}{39} (\bibinfo{year}{1981}) \bibinfo{pages}{201--225}.
  \DOIprefix\doi{10.1016/0021-9991(81)90145-5}.
%Type = Article
\bibitem[{Tryggvason et~al.(2001)Tryggvason, Brunner, Esmaeeli, Juric,
  Al-Rawahi, Tauber, Han, Nas, and Jan}]{Tryggvason2001}
\bibinfo{author}{G.~Tryggvason}, \bibinfo{author}{B.~Brunner},
  \bibinfo{author}{A.~Esmaeeli}, \bibinfo{author}{D.~Juric},
  \bibinfo{author}{N.~Al-Rawahi}, \bibinfo{author}{W.~Tauber},
  \bibinfo{author}{J.~Han}, \bibinfo{author}{S.~Nas}, \bibinfo{author}{Y.-J.
  Jan},
\newblock \bibinfo{title}{A front-tracking method for the computations of
  multiphase flow},
\newblock \bibinfo{journal}{Journal of Computational Physics}
  \bibinfo{volume}{169} (\bibinfo{year}{2001}) \bibinfo{pages}{708--759}.
  \DOIprefix\doi{10.1006/jcph.2001.6726}.
%Type = Article
\bibitem[{Sussman et~al.(1994)Sussman, Smereka, and Osher}]{Sussman1994}
\bibinfo{author}{M.~Sussman}, \bibinfo{author}{P.~Smereka},
  \bibinfo{author}{S.~Osher},
\newblock \bibinfo{title}{A level set approach for computing solutions to
  incompressible two-phase flow},
\newblock \bibinfo{journal}{Journal of Computational Physics}
  \bibinfo{volume}{114} (\bibinfo{year}{1994}) \bibinfo{pages}{146--159}.
  \DOIprefix\doi{10.1006/jcph.1994.1155}.
%Type = Incollection
\bibitem[{Fedkiw and Liu(2001)}]{Fedkiw2001}
\bibinfo{author}{R.~P. Fedkiw}, \bibinfo{author}{X.-D. Liu},
\newblock \bibinfo{title}{The ghost fluid method for viscous flows},
\newblock in: \bibinfo{booktitle}{Innovative Methods for Numerical Solution of
  Partial Differential Equations}, \bibinfo{publisher}{{WORLD} {SCIENTIFIC}},
  \bibinfo{year}{2001}, pp. \bibinfo{pages}{111--143}.
  \DOIprefix\doi{10.1142/9789812810816_0005}.
%Type = Article
\bibitem[{Brackbill et~al.(1992)Brackbill, Kothe, and Zemach}]{Brackbill1992}
\bibinfo{author}{J.~U. Brackbill}, \bibinfo{author}{D.~B. Kothe},
  \bibinfo{author}{C.~Zemach},
\newblock \bibinfo{title}{{A continuum method for modeling surface tension}},
\newblock \bibinfo{journal}{Journal of Computational Physics}
  \bibinfo{volume}{100} (\bibinfo{year}{1992}) \bibinfo{pages}{335--354}.
  \DOIprefix\doi{10.1016/0021-9991(92)90240-Y}.
%Type = Article
\bibitem[{Liu et~al.(2003)Liu, Khoo, and Yeo}]{Liu2003}
\bibinfo{author}{T.~Liu}, \bibinfo{author}{B.~Khoo}, \bibinfo{author}{K.~Yeo},
\newblock \bibinfo{title}{Ghost fluid method for strong shock impacting on
  material interface},
\newblock \bibinfo{journal}{Journal of Computational Physics}
  \bibinfo{volume}{190} (\bibinfo{year}{2003}) \bibinfo{pages}{651--681}.
  \DOIprefix\doi{10.1016/s0021-9991(03)00301-2}.
%Type = Article
\bibitem[{Merkle and Rohde(2007)}]{Merkle2007}
\bibinfo{author}{C.~Merkle}, \bibinfo{author}{C.~Rohde},
\newblock \bibinfo{title}{The sharp-interface approach for fluids with phase
  change: Riemann problems and ghost fluid techniques},
\newblock \bibinfo{journal}{{ESAIM}: Mathematical Modelling and Numerical
  Analysis} \bibinfo{volume}{41} (\bibinfo{year}{2007})
  \bibinfo{pages}{1089--1123}. \DOIprefix\doi{10.1051/m2an:2007048}.
%Type = Article
\bibitem[{Fechter et~al.(2018)Fechter, Munz, Rohde, and Zeiler}]{Fechter2018}
\bibinfo{author}{S.~Fechter}, \bibinfo{author}{C.-D. Munz},
  \bibinfo{author}{C.~Rohde}, \bibinfo{author}{C.~Zeiler},
\newblock \bibinfo{title}{Approximate {Riemann} solver for compressible liquid
  vapor flow with phase transition and surface tension},
\newblock \bibinfo{journal}{Computers {\&} Fluids} \bibinfo{volume}{169}
  (\bibinfo{year}{2018}) \bibinfo{pages}{169--185}.
  \DOIprefix\doi{10.1016/j.compfluid.2017.03.026}.
%Type = Article
\bibitem[{Gassner et~al.(2007{\natexlab{a}})Gassner, L\"orcher, and
  Munz}]{Gassner2007a}
\bibinfo{author}{G.~Gassner}, \bibinfo{author}{F.~L\"orcher},
  \bibinfo{author}{C.-D. Munz},
\newblock \bibinfo{title}{A contribution to the construction of diffusion
  fluxes for finite volume and discontinuous {G}alerkin schemes},
\newblock \bibinfo{journal}{Journal of Computational Physics}
  \bibinfo{volume}{224} (\bibinfo{year}{2007}{\natexlab{a}})
  \bibinfo{pages}{1049--1063}. \DOIprefix\doi{10.1016/j.jcp.2006.11.004}.
%Type = Article
\bibitem[{Gassner et~al.(2007{\natexlab{b}})Gassner, L\"orcher, and
  Munz}]{Gassner2007}
\bibinfo{author}{G.~Gassner}, \bibinfo{author}{F.~L\"orcher},
  \bibinfo{author}{C.-D. Munz},
\newblock \bibinfo{title}{A discontinuous {G}alerkin scheme based
  on~a~space-time expansion {II}. {V}iscous flow equations in~multi
  dimensions},
\newblock \bibinfo{journal}{Journal of Scientific Computing}
  \bibinfo{volume}{34} (\bibinfo{year}{2007}{\natexlab{b}})
  \bibinfo{pages}{260--286}. \DOIprefix\doi{10.1007/s10915-007-9169-1}.
%Type = Article
\bibitem[{L\"orcher et~al.(2008)L\"orcher, Gassner, and Munz}]{Loercher2008}
\bibinfo{author}{F.~L\"orcher}, \bibinfo{author}{G.~Gassner},
  \bibinfo{author}{C.-D. Munz},
\newblock \bibinfo{title}{An explicit discontinuous {G}alerkin scheme with
  local time-stepping for general unsteady diffusion equations},
\newblock \bibinfo{journal}{Journal of Computational Physics}
  \bibinfo{volume}{227} (\bibinfo{year}{2008}) \bibinfo{pages}{5649--5670}.
  \DOIprefix\doi{10.1016/j.jcp.2008.02.015}.
%Type = Article
\bibitem[{J\"ons et~al.(2020)J\"ons, Müller, Zeifang, and Munz}]{joens2020}
\bibinfo{author}{S.~J\"ons}, \bibinfo{author}{C.~Müller},
  \bibinfo{author}{J.~Zeifang}, \bibinfo{author}{C.-D. Munz},
\newblock \bibinfo{title}{Recent advances and complex applications of the
  compressible ghost fluid method},
\newblock \bibinfo{journal}{SEMASIMAI Springer Series, Proceedings of Numhyp
  2019. accepted. Springer,}  (\bibinfo{year}{2020}).
%Type = Article
\bibitem[{Hindenlang et~al.(2012)Hindenlang, Gassner, Altmann, Beck,
  Staudenmaier, and Munz}]{Hindenlang2012}
\bibinfo{author}{F.~Hindenlang}, \bibinfo{author}{G.~J. Gassner},
  \bibinfo{author}{C.~Altmann}, \bibinfo{author}{A.~Beck},
  \bibinfo{author}{M.~Staudenmaier}, \bibinfo{author}{C.-D. Munz},
\newblock \bibinfo{title}{Explicit discontinuous {Galerkin} methods for
  unsteady problems},
\newblock \bibinfo{journal}{Computers {\&} Fluids} \bibinfo{volume}{61}
  (\bibinfo{year}{2012}) \bibinfo{pages}{86--93}.
  \DOIprefix\doi{10.1016/j.compfluid.2012.03.006}.
%Type = Article
\bibitem[{Krais et~al.(2020)Krais, Beck, Bolemann, Frank, Flad, Gassner,
  Hindenlang, Hoffmann, Kuhn, Sonntag, and Munz}]{Krais2020}
\bibinfo{author}{N.~Krais}, \bibinfo{author}{A.~Beck},
  \bibinfo{author}{T.~Bolemann}, \bibinfo{author}{H.~Frank},
  \bibinfo{author}{D.~Flad}, \bibinfo{author}{G.~Gassner},
  \bibinfo{author}{F.~Hindenlang}, \bibinfo{author}{M.~Hoffmann},
  \bibinfo{author}{T.~Kuhn}, \bibinfo{author}{M.~Sonntag},
  \bibinfo{author}{C.-D. Munz},
\newblock \bibinfo{title}{{FLEXI}: A high order discontinuous {Galerkin}
  framework for hyperbolic{\textendash}parabolic conservation laws},
\newblock \bibinfo{journal}{Computers {\&} Mathematics with Applications}
  (\bibinfo{year}{2020}). \DOIprefix\doi{10.1016/j.camwa.2020.05.004}.
%Type = Article
\bibitem[{Sonntag and Munz(2016)}]{Sonntag2016}
\bibinfo{author}{M.~Sonntag}, \bibinfo{author}{C.-D. Munz},
\newblock \bibinfo{title}{Efficient parallelization of a shock capturing for
  discontinuous {Galerkin} methods using finite volume sub-cells},
\newblock \bibinfo{journal}{Journal of Scientific Computing}
  \bibinfo{volume}{70} (\bibinfo{year}{2016}) \bibinfo{pages}{1262--1289}.
  \DOIprefix\doi{10.1007/s10915-016-0287-5}.
%Type = Article
\bibitem[{Bassi and Rebay(1997)}]{Bassi1997}
\bibinfo{author}{F.~Bassi}, \bibinfo{author}{S.~Rebay},
\newblock \bibinfo{title}{A high-order accurate discontinuous finite element
  method for the numerical solution of the compressible
  navier{\textendash}stokes equations},
\newblock \bibinfo{journal}{Journal of Computational Physics}
  \bibinfo{volume}{131} (\bibinfo{year}{1997}) \bibinfo{pages}{267--279}.
  \DOIprefix\doi{10.1006/jcph.1996.5572}.
%Type = Article
\bibitem[{Kennedy and Carpenter(2003)}]{Kennedy2003}
\bibinfo{author}{C.~A. Kennedy}, \bibinfo{author}{M.~H. Carpenter},
\newblock \bibinfo{title}{Additive runge{\textendash}kutta schemes for
  convection{\textendash}diffusion{\textendash}reaction equations},
\newblock \bibinfo{journal}{Applied Numerical Mathematics} \bibinfo{volume}{44}
  (\bibinfo{year}{2003}) \bibinfo{pages}{139--181}.
  \DOIprefix\doi{10.1016/s0168-9274(02)00138-1}.
%Type = Article
\bibitem[{Dumbser and Loub{\`{e}}re(2016)}]{Dumbser2016a}
\bibinfo{author}{M.~Dumbser}, \bibinfo{author}{R.~Loub{\`{e}}re},
\newblock \bibinfo{title}{A simple robust and accurate a posteriori sub-cell
  finite volume limiter for the discontinuous galerkin method on unstructured
  meshes},
\newblock \bibinfo{journal}{Journal of Computational Physics}
  \bibinfo{volume}{319} (\bibinfo{year}{2016}) \bibinfo{pages}{163--199}.
  \DOIprefix\doi{10.1016/j.jcp.2016.05.002}.
%Type = Phdthesis
\bibitem[{Fechter(2015)}]{Fechter2015a}
\bibinfo{author}{S.~Fechter}, \bibinfo{title}{Compressible Multi-Phase
  Simulation at Extreme Conditions Using a Discontinuous {Galerkin} Scheme},
  Ph.D. thesis, \bibinfo{year}{2015}. \DOIprefix\doi{10.18419/opus-3982}.
%Type = Article
\bibitem[{Peng et~al.(1999)Peng, Merriman, Osher, Zhao, and Kang}]{Peng1999}
\bibinfo{author}{D.~Peng}, \bibinfo{author}{B.~Merriman},
  \bibinfo{author}{S.~Osher}, \bibinfo{author}{H.~Zhao},
  \bibinfo{author}{M.~Kang},
\newblock \bibinfo{title}{A {PDE}-based fast local level set method},
\newblock \bibinfo{journal}{Journal of Computational Physics}
  \bibinfo{volume}{155} (\bibinfo{year}{1999}) \bibinfo{pages}{410--438}.
  \DOIprefix\doi{10.1006/jcph.1999.6345}.
%Type = Article
\bibitem[{Xu and Liu(2011)}]{Xu2011}
\bibinfo{author}{L.~Xu}, \bibinfo{author}{T.~Liu},
\newblock \bibinfo{title}{{Accuracies and conservation errors of various ghost
  fluid methods for multi-medium Riemann problem}},
\newblock \bibinfo{journal}{Journal of Computational Physics}
  \bibinfo{volume}{230} (\bibinfo{year}{2011}) \bibinfo{pages}{4975--4990}.
  \DOIprefix\doi{10.1016/j.jcp.2011.03.021}.
%Type = Article
\bibitem[{Xu et~al.(2016)Xu, Feng, and Liu}]{Xu2016}
\bibinfo{author}{L.~Xu}, \bibinfo{author}{C.~Feng}, \bibinfo{author}{T.~Liu},
\newblock \bibinfo{title}{{Practical Techniques in Ghost Fluid Method for
  Compressible Multi-Medium Flows}},
\newblock \bibinfo{journal}{Commun. Comput. Phys.} \bibinfo{volume}{20}
  (\bibinfo{year}{2016}) \bibinfo{pages}{619--659}.
  \DOIprefix\doi{10.4208/cicp.190315.290316a}.
%Type = Article
\bibitem[{Fechter and Munz(2015)}]{Fechter2015}
\bibinfo{author}{S.~Fechter}, \bibinfo{author}{C.-D. Munz},
\newblock \bibinfo{title}{A discontinuous {Galerkin}-based sharp-interface
  method to simulate three-dimensional compressible two-phase flow},
\newblock \bibinfo{journal}{International Journal for Numerical Methods in
  Fluids} \bibinfo{volume}{78} (\bibinfo{year}{2015})
  \bibinfo{pages}{413--435}. \DOIprefix\doi{10.1002/fld.4022}.
%Type = Article
\bibitem[{F{\"o}ll et~al.(2020)F{\"o}ll, M{\"u}ller, Zeifang, and
  Munz}]{foll2020novel}
\bibinfo{author}{F.~F{\"o}ll}, \bibinfo{author}{C.~M{\"u}ller},
  \bibinfo{author}{J.~Zeifang}, \bibinfo{author}{C.-D. Munz},
\newblock \bibinfo{title}{A novel regularization strategy for the local
  discontinuous galerkin method for level-set reinitialization},
\newblock \bibinfo{journal}{arXiv preprint arXiv:2007.06883}
  (\bibinfo{year}{2020}).
%Type = Article
\bibitem[{Toro et~al.(1994)Toro, Spruce, and Speares}]{Toro1994}
\bibinfo{author}{E.~F. Toro}, \bibinfo{author}{M.~Spruce},
  \bibinfo{author}{W.~Speares},
\newblock \bibinfo{title}{Restoration of the contact surface in the
  {HLL}-{Riemann} solver},
\newblock \bibinfo{journal}{Shock Waves} \bibinfo{volume}{4}
  (\bibinfo{year}{1994}) \bibinfo{pages}{25--34}.
  \DOIprefix\doi{10.1007/bf01414629}.
%Type = Article
\bibitem[{Hu et~al.(2009)Hu, Adams, and Iaccarino}]{Hu2009}
\bibinfo{author}{X.~Hu}, \bibinfo{author}{N.~Adams},
  \bibinfo{author}{G.~Iaccarino},
\newblock \bibinfo{title}{{On the {HLLC} Riemann Solver for Interface
  Interaction in Compressible Multi-Fluid Flow}},
\newblock \bibinfo{journal}{Journal of Computational Physics}
  \bibinfo{volume}{228} (\bibinfo{year}{2009}) \bibinfo{pages}{6572--6589}.
  \DOIprefix\doi{10.1016/j.jcp.2009.06.002}.
%Type = Article
\bibitem[{Jaegle et~al.(2012)Jaegle, Rohde, and Zeiler}]{Jaegle2012}
\bibinfo{author}{F.~Jaegle}, \bibinfo{author}{C.~Rohde},
  \bibinfo{author}{C.~Zeiler},
\newblock \bibinfo{title}{A multiscale method for compressible liquid-vapor
  flow with surface tension},
\newblock \bibinfo{journal}{{ESAIM}: Proceedings} \bibinfo{volume}{38}
  (\bibinfo{year}{2012}) \bibinfo{pages}{387--408}.
  \DOIprefix\doi{10.1051/proc/201238022}.
%Type = Article
\bibitem[{Hitz et~al.(2021)Hitz, J\"ons, Heinen, Vrabec, and Munz}]{Hitz2020}
\bibinfo{author}{T.~Hitz}, \bibinfo{author}{S.~J\"ons},
  \bibinfo{author}{M.~Heinen}, \bibinfo{author}{J.~Vrabec},
  \bibinfo{author}{C.-D. Munz},
\newblock \bibinfo{title}{Comparison of macro- and microscopic solutions of the
  {Riemann} problem {II}. {T}wo-phase shock tube},
\newblock \bibinfo{journal}{Journal of Computational Physics}
  \bibinfo{volume}{429} (\bibinfo{year}{2021}) \bibinfo{pages}{110027}.
  \DOIprefix\doi{10.1016/j.jcp.2020.110027}.
%Type = Article
\bibitem[{J{\"o}ns and Munz(2023)}]{joens2023}
\bibinfo{author}{S.~J{\"o}ns}, \bibinfo{author}{C.-D. Munz},
\newblock \bibinfo{title}{Riemann solvers for phase transition in a
  compressible sharp-interface method},
\newblock \bibinfo{journal}{Applied Mathematics and Computation}
  \bibinfo{volume}{440} (\bibinfo{year}{2023}) \bibinfo{pages}{127624}.
%Type = Article
\bibitem[{M{\"u}ller et~al.(2023)M{\"u}ller, Mossier, and Munz}]{muller2023}
\bibinfo{author}{C.~M{\"u}ller}, \bibinfo{author}{P.~Mossier},
  \bibinfo{author}{C.-D. Munz},
\newblock \bibinfo{title}{A sharp interface framework based on the inviscid
  godunov-peshkov-romenski equations: Simulation of evaporating fluids},
\newblock \bibinfo{journal}{Journal of Computational Physics}
  \bibinfo{volume}{473} (\bibinfo{year}{2023}) \bibinfo{pages}{111737}.
%Type = Article
\bibitem[{Zeifang and Beck(2021)}]{zeifang2021low}
\bibinfo{author}{J.~Zeifang}, \bibinfo{author}{A.~Beck},
\newblock \bibinfo{title}{A low mach number imex flux splitting for the level
  set ghost fluid method},
\newblock \bibinfo{journal}{Communications on Applied Mathematics and
  Computation}  (\bibinfo{year}{2021}) \bibinfo{pages}{1--29}.
%Type = Book
\bibitem[{Schlichting and Gersten(2017)}]{Schlichting2017a}
\bibinfo{author}{H.~Schlichting}, \bibinfo{author}{K.~Gersten},
  \bibinfo{title}{{Boundary-Layer Theory}}, volume~\bibinfo{volume}{7},
  \bibinfo{publisher}{Springer Berlin Heidelberg}, \bibinfo{address}{Berlin,
  Heidelberg}, \bibinfo{year}{2017}. \DOIprefix\doi{10.1007/978-3-662-52919-5}.
%Type = Incollection
\bibitem[{Dumbser et~al.(2018)Dumbser, Peshkov, and Romenski}]{Dumbser2018}
\bibinfo{author}{M.~Dumbser}, \bibinfo{author}{I.~Peshkov},
  \bibinfo{author}{E.~Romenski},
\newblock \bibinfo{title}{A unified hyperbolic formulation for viscous fluids
  and elastoplastic solids},
\newblock in: \bibinfo{booktitle}{Theory, Numerics and Applications of
  Hyperbolic Problems {II}}, \bibinfo{publisher}{Springer International
  Publishing}, \bibinfo{year}{2018}, pp. \bibinfo{pages}{451--463}.
  \DOIprefix\doi{10.1007/978-3-319-91548-7_34}.
%Type = Article
\bibitem[{Lalanne et~al.(2015)Lalanne, {Rueda Villegas}, Tanguy, Risso, and
  Villegas}]{Lalanne2015a}
\bibinfo{author}{B.~Lalanne}, \bibinfo{author}{L.~{Rueda Villegas}},
  \bibinfo{author}{S.~Tanguy}, \bibinfo{author}{F.~Risso},
  \bibinfo{author}{R.~Villegas},
\newblock \bibinfo{title}{{On the computation of viscous terms for
  incompressible two-phase flows with Level Set/Ghost Fluid Method On the
  computation of viscous terms for incompressible two-phase flows with Level
  Set/Ghost Fluid Method Open Archive TOULOUSE Archive Ouverte (OATAO)}},
\newblock \bibinfo{journal}{Journal of Computational Physics}
  \bibinfo{volume}{301} (\bibinfo{year}{2015}) \bibinfo{pages}{289--307}.
  \DOIprefix\doi{10.1016/j.jcp.2015.08.036ï}.
%Type = Article
\bibitem[{Prosperetti(1980)}]{Prosperetti1980}
\bibinfo{author}{A.~Prosperetti},
\newblock \bibinfo{title}{Normal-mode analysis for the oscillations of a
  viscous liquid drop in an immiscible liquid},
\newblock \bibinfo{journal}{J. Mecanique} \bibinfo{volume}{19}
  (\bibinfo{year}{1980}) \bibinfo{pages}{149--182}.
%Type = Book
\bibitem[{Lamb(1924)}]{lamb1924hydrodynamics}
\bibinfo{author}{H.~Lamb}, \bibinfo{title}{Hydrodynamics},
  \bibinfo{publisher}{University Press}, \bibinfo{year}{1924}.
%Type = Inproceedings
\bibitem[{Winter et~al.(2019)Winter, Kaiser, Adami, and
  Adams}]{winter2019numerical}
\bibinfo{author}{J.~Winter}, \bibinfo{author}{J.~Kaiser},
  \bibinfo{author}{S.~Adami}, \bibinfo{author}{N.~Adams},
\newblock \bibinfo{title}{Numerical investigation of 3d drop-breakup mechanisms
  using a sharp interface level-set method},
\newblock in: \bibinfo{booktitle}{11th International Symposium on Turbulence
  and Shear Flow Phenomena, TSFP 2019}, \bibinfo{year}{2019}.

\end{thebibliography}

\end{document}